\documentclass[aps,preprint,nofootinbib,floatfix,a4paper]{revtex4-1}
\pdfoutput=1
\usepackage{hyperref}

\usepackage{graphicx}
\usepackage[latin1]{inputenc}
\usepackage{amsmath,amssymb}
\usepackage{slashed}
\usepackage{epstopdf}
\usepackage{geometry}    

\usepackage{feynmp-auto}
\usepackage{pgfplots}
\usepackage{subfig}
\usepackage{tabularx}
\newcolumntype{C}{>{\centering\arraybackslash}X} 
\def\be{\begin{equation}}
\def\ee{\end{equation}}
\def\bea{\begin{eqnarray}}
\def\eea{\end{eqnarray}}

\begin{document}

\title{Top Quark Rare Decays  \\
via Loop-Induced FCNC Interactions in\\
Extended Mirror Fermion Model}

\author{P. Q. Hung}
\email[]{pqh@virginia.edu}
\affiliation{Department of Physics, University of Virginia, Charlottesville, VA 22904-4714, USA\\
and\\
Center for Theoretical and Computational Physics, Hue University College of Education, Hue, Vietnam}

\author{Yu-Xiang Lin}
\email[]{x90417@gmail.com}
\affiliation{Department of Physics, National Taiwan Normal University, Taipei 116, Taiwan}

\author{Chrisna Setyo Nugroho}
\email[]{setyo13nugros@gmail.com}
\affiliation{Department of Physics, National Taiwan Normal University, Taipei 116, Taiwan}

\author{Tzu-Chiang Yuan}
\email[]{tcyuan@phys.sinica.edu.tw}
\affiliation{Institute of Physics,\\ Academia Sinica,\\ Nangang, Taipei 11529, Taiwan\\
and\\
Physics Division,\\ National Center for Theoretical Sciences, Hsinchu, Taiwan}

\date{\today}                                          

\begin{abstract}

Flavor changing neutral current (FCNC) interactions for a top quark $t$ decays into
$Xq$ with $X$ represents a neutral gauge or Higgs boson,
and $q$ a up- or charm-quark are highly suppressed in the Standard Model (SM)
due to the Glashow-Iliopoulos-Miami mechanism. 
Whilst current limits on the branching ratios of these processes 
have been established at the order of $10^{-4}$ from
the Large Hadron Collider experiments, 
SM predictions are at least nine orders of magnitude below.
In this work, we study some of these FCNC processes in the context 
of an extended mirror fermion model, originally proposed to implement 
the electroweak scale seesaw mechanism for non-sterile right-handed neutrinos. 
We show that one can probe the process 
$t \to Zc$ for a wide range of parameter space 
with branching ratios varying from $10^{-6}$ to $10^{-8}$, 
comparable with various new physics models including the general two Higgs doublet model
with or without flavor violations at tree level, 
minimal supersymmetric standard model with or without $R$-parity, 
and extra dimension model.

\end{abstract}

\maketitle

\section{Introduction}

Absence of flavor changing neutral current (FCNC) interactions in the Standard Model (SM) 
at tree level is quite a unique property due to the special quantum numbers of 
the three generations of fermions (quarks and leptons) 
and one Higgs doublet under the gauge group of
$SU(3)_C \times SU(2)_L \times U(1)_Y$. FCNC interactions can nevertheless be induced at the quantum 
loop level and therefore are suppressed by the GIM mechanism~\cite{Glashow:1970gm}.
Experimental results for various FCNC processes in the kaon, $D$ and $B$ meson systems are all in line
with the SM expectations. 

For the heavy top quark $t$, the story is quite different.
Since there is no time for the heavy top quark to form bound states, 
we can discuss its free decay, 
like the dominant decay mode $t \to W^+ b$ at tree level or its rare FCNC decays. 
The SM branching ratios ${\mathcal B}(t \to X q)$ where $X$ denotes 
one of the following neutral particle $Z, \gamma, g$ or $h$ in SM, and $q$ denotes the light 
$u$ or $c$ quark, vary in the range $10^{-17} -10^{-12}$~\cite{AguilarSaavedra:2004wm}, which are unobservable at the present technology.
However, in many models beyond the SM, branching ratios for some of these processes of order up to $10^{-3}$  can be achieved. Observations of these rare top quark FCNC decays 
at the Large Hadron Collider (LHC) with significant larger branching ratios than the SM predictions would then be clear signals, albeit  indirect, of new physics. 

Indeed LHC can be considered as a top quark factory, 
estimated to produce $10^8$ $t \bar t$ pair with an integrated luminosity of 100 inverse femtobarn.
For an updated review on top quark properties at the LHC, see~\cite{Soares-Moriond}.
The current limits for $t \to Zq$~\cite{CMStZq,CMS-PAS-14-20,Aad:2015uza,ATLAS:2017beb} 
are
\begin{equation}
{\mathcal B}(t \to Z u)   \leq  \left\{
\begin{array}{l}
2.2 \times 10^{-4} \;  [{\rm CMS}] \; , \\
1.7 \times 10^{-4}  \; [{\rm ATLAS}] \; ,
 \end{array}
\right.
\end{equation}
\begin{equation}
{\mathcal B}(t \to Z c)  \leq \left\{
\begin{array}{l}
4.9 \times 10^{-4}  \; [{\rm CMS}] \; , \\
2.3 \times 10^{-4}  \; [{\rm ATLAS}] \; ;
 \end{array} 
 \right.
\end{equation}
and for $t \to \gamma q$, we have the limits from CMS~\cite{Khachatryan:2015att}
\begin{equation}
\begin{aligned}
{\mathcal B}(t \to \gamma u)   & \leq
 1.3 \times 10^{-4} \; , \\
{\mathcal B}(t \to \gamma c)  & \leq
 1.7 \times 10^{-3}  \; .
\end{aligned}
\end{equation}
Projected limits for the above as well as other FCNC processes for the top quark 
are expected to be improved constantly in the future 
at the LHC.
Thus searching for or discovery of any one of these FCNC rare top decays $t \to X q$ 
at the LHC would be providing
interesting constraints or discriminations  of various new physics models in the future.

Over the years FCNC top quark decays had been studied intensively in the literature for many new physics 
models, like the minimal supersymmetric standard model (MSSM) with~\cite{Cao:2007dk} or 
without~\cite{Yang:1997dk,Eilam:2001dh} $R$-parity, 
flavor conserving~\cite{Eilam:1990zc} or flavor violating~\cite{,Atwood:1996vj,Bejar:2006ww,Kao:2011aa,Chen:2013qta} two Higgs doublet model (2HDM),
aligned two Higgs doublet model (A2HDM)~\cite{Abbas:2015cua},
warped extra-dimensions~\cite{Agashe:2006wa,Agashe:2009di},
and effective Lagrangian framework~\cite{Hesari:2015oya}, {\it etc}.
Branching ratios for FCNC top quark decays in all these models are typically 
many orders of magnitude above the SM and 
some of them may lead to detectable signals at the LHC.

In this work, we compute the FCNC decays of $t \to Vq$ $(V=Z,\gamma; \, q=u,c)$ 
in an extension of mirror fermion model~\cite{Hung:2015nva} originally 
proposed by one of us~\cite{Hung:2006ap}. 
In contrast with various left-right symmetric models, the model in~\cite{Hung:2006ap}
did not include the gauge group $SU(2)_R$
while adding the mirror partners of the SM fermions.
Despite having the same SM gauge group, the scalar sector must be enlarged. 
In additional to employ the bi-triplets in the Georgi-Machacek (GM) 
model~\cite{Georgi:1985nv,Chanowitz:1985ug} 
and a Higgs singlet to implement the electroweak scale seesaw 
mechanism for the non-sterile right-handed neutrino masses~\cite{Hung:2006ap}, one needs 
to add a mirror Higgs doublet~\cite{Hoang:2014pda} 
in the scalar sector so as to make consistency with the various signal strengths 
of the 125 GeV Higgs measured at the LHC.
We will briefly review this class of mirror fermion model and its further extension with 
a horizontal $A_4$ symmetry in the following section.

We layout the paper as follows. In Section~\ref{model}, 
after giving a brief highlight on some of the salient features of the model, we 
present the relevant interaction Lagrangian.
Our calculation and analysis are presented in Section~\ref{calculation} and~\ref{analysis} respectively. We finally summarize our results in Section~\ref{summary}. 
Analytical expressions for the loop functions are collected in the Appendix.

\section{The Mirror Fermion Model}
\label{model}

\subsection{A Lightning Review}

As already eluded to in the Introduction, the mirror fermion model in~\cite{Hung:2006ap} was devised to implement the so-called electroweak scale seesaw mechanism for the neutrino masses.  We first list 
the particle content of the model in Table~\ref{particlecontent} for further discussions.
One special feature of this mirror model is to treat the right-handed neutrino in each generation to be non-sterile 
by grouping it with a new heavy mirror right-handed charged lepton 
into a weak doublet $l^M_{Ri}$, regarded as the mirror of the SM doublet $l_{Li}$ with $i$ labelling the generation.
When the Higgs singlet $\phi_{0S}$ develops a small vacuum expectation value (VEV) of order $10^5$ eV, 
through its Yukawa couplings between the SM lepton doublets and their mirror partners, 
it can provide a small Dirac mass term for the neutrinos.
On the other hand, when the triplet $\tilde \chi$ field with hypercharge $Y/2 = 1$ in Table~\ref{particlecontent}
develops a VEV of order $v_{\rm SM} = 246 $ GeV, 
a Majorana mass term of electroweak scale can be generated through its
Yukawa couplings among these new mirror lepton doublets.
Details of this electroweak scale seesaw mechanism in the mirror fermion model 
can be found in~\cite{Hung:2006ap}.

\begin{table}
\begin{tabular}{|c|c|}
\hline
Fields & ($SU(3)\, , \, SU(2) \, , \,  U(1)_Y$ ; $A_4$) \\ 
\hline
\hline
$ l_{Li}= \left( \begin{array}{c} \nu_L \\ e_L \end{array} \right)_i$ \, , \,  $l^M_{Ri}= \left( \begin{array}{c} \nu_R \\ e^M_R \end{array} \right)_i$ & (1, 2, $-\frac{1}{2}$ ; 3)\\
$e_{Ri}$ \, , \, $e^M_{Li}$ &  (1, 1, $-1$ ; 3) \\
\hline
$ q_{Li}= \left( \begin{array}{c} u_L \\ d_L \end{array} \right)_i$ \, , \,  $q^M_{Ri}= \left( \begin{array}{c} u^M_R \\ d^M_R \end{array} \right)_i$ & (3, 2, $\frac{1}{6}$ ; 3)\\
$u_{Ri}$ \, , \, $u^M_{Li}$ &  (3, 1, $\frac{2}{3}$ ; 3) \\
$d_{Ri}$ \, , \, $d^M_{Li}$ &  (3, 1, $-\frac{1}{3}$ ; 3) \\
\hline
\hline
$\phi_{0S}$ &  (1, 1, 0 ; 1) \\
$\phi_{iS}$ & (1, 1, 0 ; 3) \\
\hline
$\Phi$ \, , \, $\Phi_M$ &  (1, 2, $\frac{1}{2}$ ; 1) \\
\hline
$\, \xi \, $ &  (1, 3, 0 ; 1)\\
$\, \tilde \chi \, $ & (1, 3, 1 ; 1)\\
\hline
\end{tabular}
\caption{The Standard Model quantum numbers of the fermion and scalar sectors in the extended mirror model together with their assignments under the horizontal $A_4$ symmetry.}
\label{particlecontent}
\end{table} 

Note that the other triplet $\xi$ which has zero hypercharge 
is grouped with  $\tilde \chi$ to form the bi-triplets in the GM model 
to maintain the custodial symmetry and therefore 
the $\rho$ parameter equals unity at tree level. 
In~\cite{Hoang:2013jfa}, the potential dangerous contributions from the GM triplets to the $S$ and $T$ oblique parameters are shown to be partially cancelled by the opposite contributions from mirror fermions such that the model is still healthy against electroweak precision tests.

Mirrors of other SM fermions, both leptons and quarks, 
can be introduced in the same way as listed in Table~\ref{particlecontent}.
Searches for these heavy mirror fermions at the LHC were presented
in~\cite{Chakdar:2015sra} and~\cite{Chakdar:2016adj} for the mirror quarks
and mirror leptons respectively.

In order to reproduce the signal strengths of $h_{125} \to \gamma\gamma$ and $h_{125} \to Z \gamma$ for the 125 GeV Higgs observed at the LHC, a mirror Higgs doublet  $\Phi_M$ of the
SM one $\Phi$ was introduced~\cite{Hoang:2014pda}. 
We note that mixing effects among the two doublets as well as with 
the triplet $\xi$ must be taken into account in order to satisfy the LHC results.
A global $U(1) \times U(1)$ symmetry was also enforced in the 
Yukawa interactions so that the SM Higgs doublet only couples to the SM fermions and the mirror Higgs doublet only couples to the mirror fermions.
Thus there is no FCNC Higgs interactions at tree level in the model. Processes like 
$h \to \tau \mu$~\cite{Chang:2016ave} and $t \to h q$ can only occur at the quantum loop level. 

To address the issues of neutrino and charged lepton masses and mixings, 
the original mirror model was extended in~\cite{Hung:2015nva} by introducing
a horizontal family symmetry of the tetrahedral group $A_4$. 
The $A_4$ assignments of all the scalars and fermions as well as their SM quantum numbers are shown at the last column in Table~\ref{particlecontent}. 
The lone singlet $\phi_{0S}$ is now accompanied with a $A_4$ triplet $\vec\phi_{S}=(\phi_{1S},\phi_{2S},\phi_{3S})$. 
Both $\phi_{0S}$ and $\vec \phi_{S}$ are electroweak singlets and they are the 
only fields communicating the SM sector with the mirror sector 
through the Yukawa couplings, which must be invariant under both gauge symmetry and $A_4$. Other scalars are $A_4$ singlets.

Phenomenological implications of the 
extended mirror fermion model with the $A_4$ symmetry 
have been explored for the charged lepton flavor violating (CLFV) processes 
$\mu \to e \gamma$~\cite{Hung:2015hra}, 
$\mu - e$ conversion~\cite{Hung:2017voe} and 
$h_{125} \to \tau \mu$~\cite{Chang:2016ave},
as well as for the electron electric dipole moment~\cite{Chang:2017vzi}.
Here we will explore its implication in the rare FCNC top decays.
Implications of the $A_4$ symmetry for the quark masses and mixings 
will be given in~\cite{quarkA4}.

\subsection{Interaction Lagrangian for Quarks and Their Mirrors}

Here we will write down the interactions for the quarks and their mirrors that are relevant to the 
FCNC processes $t \to Vq$ that we are studying. Since the result for 
$t \to gq$ can be easily obtained from that of $t \to \gamma q$, we will not present detailed formulas for the former process.
As for $t \to h q$, one must consider the mixing effects from the more complicated Higgs sector in the extended mirror model. We will leave it for future work.

\subsubsection{Quark Yukawa Couplings with $A_4$ Symmetry}

Recall that the tetrahedron symmetry group $A_4$ has four irreducible 
representations $\bf 1$, $\bf 1'$, $\bf 1''$, 
and $\bf 3$ with the following multiplication rule
\begin{eqnarray}
{\bf 3} \times {\bf 3} & = &   {\bf 3_1} (23, 31, 12) + {\bf 3_2} (32, 13, 21) \nonumber \\
&+& {\bf 1} (11+ 22 + 33) + {\bf 1'} (11 + \omega^2 22 + \omega 33) 
+ {\bf 1''} (11 + \omega 22 + \omega^2 33)
\label{A4rules}
\end{eqnarray}
where $\omega = e^{2 \pi i/3} = -\frac{1}{2} + i \frac{\sqrt 3}{2}$.

Using the above $A_4$ multiplication rules one can construct new Yukawa couplings in the leptonic sector, which are
both gauge invariant and $A_4$ symmetric, to implement small Dirac neutrino masses in electroweak seesaw
and to discuss charged lepton mixings~\cite{Hung:2015nva}.

In the same vein, one can write down the following new Yukawa couplings 
for the quarks and their mirrors (both in the flavor basis with subscripts ``0")
with the scalar singlets, which are both gauge invariant and $A_4$ symmetric,
\bea
- {\cal L}_Y 
& \supset & g^Q_{0S} \phi_{0S} (\overline{q_{L,0}}   q^{M}_{R,0})_{\bf 1} 
+ g^Q_{1S} \vec \phi_S \cdot (\overline{q_{L,0}} \times q_{R,0}^{M})_{\bf 3_1}
+ g^Q_{2S} \vec \phi_S \cdot (\overline{q_{L,0}} \times q_{R,0}^{M})_{\bf 3_2} 
\nonumber \\
&+&  g^{u}_{0S} \phi_{0S} (\overline{u_{R,0}}   u^{M}_{L,0})_{\bf 1} 
+ g^{u}_{1S} \vec \phi_S \cdot (\overline{u_{R,0}} \times u_{L,0}^{M})_{\bf 3_1}
+ g^{u}_{2S} \vec \phi_S \cdot (\overline{u_{R,0}} \times u_{L,0}^{M})_{\bf 3_2} 
 \\
&+&  g^{d}_{0S} \phi_{0S} (\overline{d_{R,0}}   d^{M}_{L,0})_{\bf 1} 
+ g^{d}_{1S} \vec \phi_S \cdot (\overline{d_{R,0}} \times d_{L,0}^{M})_{\bf 3_1}
+ g^{d}_{2S} \vec \phi_S \cdot (\overline{d_{R,0}} \times d_{L,0}^{M})_{\bf 3_2} 
+ {\rm H.c.} \nonumber
\label{A4Lang1}
\eea
where $g^{Q,u,d}_{0S}$, $g^{Q,u,d}_{1S}$ and $g^{Q,u,d}_{2S}$ are in general complex
coupling constants. Implications of the above Yukawa interactions on the quark mixings will be presented in~\cite{quarkA4}.

Next we move to the physical basis by making the following unitary transformations
on the left-handed fields
$$
u_{L,0} =  V^u_L u_L,  \;
d_{L,0}  =  V^d_L d_L,  \;
u^M_{L,0}  =  V^{u^M}_L u^M_L, \;
d^M_{L,0}  =  V^{d^M}_L d^M_L,
$$
and similarly for the right-handed fields
$$
u_{R,0} =  V^u_R u_R,  \;
d_{R,0}  =  V^d_R d_R,  \;
u^M_{R,0}  =  V^{u^M}_R u^M_R, \;
d^M_{R,0}  =  V^{d^M}_R d^M_R.
$$
We can then recast the Yukawa interactions in the following form
\begin{equation}
    \begin{aligned}
        { \mathcal{L} }_{Y} \supset &-\bar { u } \left({ { V }_{ L }^{ u } }^{ \dagger  }{ M }_{ S }^{ Q }(\phi){ V }_{ R }^{ u^M }P_R+{ { V }_{ R }^{ u } }^{ \dagger  }{ M }_{ S }^{ u }(\phi){ V }_{ L }^{ u^M }P_L\right){ u }^{ M }\\
        &-\bar { d } \left({ { V }_{ L }^{ d } }^{ \dagger  }{ M }_{ S }^{ Q }(\phi){ V }_{ R }^{ d^M }P_R+{ { V }_{ R }^{ d } }^{ \dagger  }{ M }_{ S }^{ d }(\phi){ V }_{ L }^{ d^M }P_L\right){ d }^{ M }+{\rm H.c.}
    \end{aligned}
\end{equation}
with $P_{L,R}=(1\mp\gamma_5)/2$.
Here $M_S^{Q,u,d}(\phi)$ are three field-dependent three by three matrices which can be decomposed 
in terms of the four scalar fields according to
\begin{equation}
    { M }_{ S }^{ Q }(\phi)={ M }^{ Q,0 }{ \phi  }_{ 0S }+{ M }^{ Q,1 }{ \phi  }_{ 1S }+{ M }^{ Q,2 }{ \phi  }_{ 2S }+{ { M } }^{ Q,3 }{ \phi  }_{ 3S }\;,
\end{equation}
where
\begin{equation}
    \begin{aligned}
        &{ M }^{ Q,0 }=\begin{pmatrix}  { g }_{ 0S }^{ Q }  & 0 & 0 \\ 0 & { g }_{ 0S }^{ Q } & 0 \\ 0 & 0 & { g }_{ 0S }^{ Q } \end{pmatrix} \quad , \quad { M }^{ Q,1 }=\begin{pmatrix} 0 & 0 & 0 \\ 0 & 0 &  { g }_{ 1S }^{ Q }  \\ 0 & { g }_{ 2S }^{ Q } & 0 \end{pmatrix} \; ,\\
        &{ M }^{ Q,2 }=\begin{pmatrix} 0 & 0 &  { g }_{ 2S }^{ Q }  \\ 0 & 0 & 0 \\  { g }_{ 1S }^{ Q }  & 0 & 0 \end{pmatrix}\quad , \quad { M }^{ Q,3 }=\begin{pmatrix} 0 &  { g }_{ 1S }^{ Q }  & 0 \\  { g }_{ 2S }^{ Q }  & 0 & 0 \\ 0 & 0 & 0 \end{pmatrix} \; ,
    \end{aligned}
    \label{MQk}
\end{equation}
and similar decompositions for $M^{u}_S(\phi)$ and $M^{d}_S(\phi)$ with $M^{u,k}$ and $M^{d,k}$
obtained by the substitutions of $g_{iS}^Q \to$ $g_{iS}^u$ and $g_{iS}^d$ respectively in Eq.~(\ref{MQk}).
Introducing the following combinations of the coupling matrices 
${ M }^{ Q,k }$ and ${ M }^{ q,k }$ $(k=0,1,2,3)$ with the fermion mixing matrices 
\begin{equation}
\begin{aligned}
{V^{q,k}_L}  & \equiv { { V }_{ L }^{ q } }^{ \dagger  }{ M }^{ Q,k }{ V }_{ R }^{ q^M }\;, \\
{V^{q,k}_R} & \equiv { { V }_{ R }^{ q } }^{ \dagger  }{ M }^{ q,k }{ V }_{ L }^{ q^M }\;,
\label{VqkLR}
\end{aligned}
\end{equation} 
we arrive at the final form of the Yukawa interactions
 \begin{equation}
 { \mathcal{L} }_{ Y } \supset -{ \sum _{ k=0 }^{ 3 }{ \sum _{ i,j=1 }^{ 3 }{ \bar { { u} }_{ i} \left\{ { { V }_{ L }^{ u,k } }_{ ij }P_R+{ { V }_{ R }^{ u,k } }_{ ij }P_L \right\} { u }_{ j }^{ M }{ \phi  }_{ kS } + (u \leftrightarrow d)
 +{\rm H.c.} }  }  } 
\end{equation}
We have combined the four scalars $\phi_{0S}$ and $\vec \phi_{S}$ into $\phi_{kS}$ with $k=0,1,2,3$.
For the FCNC rare top decays that we are studying, only ${ { V }_{ (L,R)  }^{ u,k } }$ are relevant.


\subsubsection{Neutral Currents}

We also need the neutral current interactions for the SM $Z$ boson and 
photon couple to the quarks and their mirrors.

\begin{equation}
{\mathcal L}_{\rm NC} \supset 
g Z^\mu J_\mu^Z + e A^\mu J^{\rm EM}_\mu 
\end{equation}
with 
\begin{equation}
\begin{aligned}
J_\mu^Z = \frac{1}{\cos \theta_W} &
\left[ 
\bar q_L \gamma_\mu \left( T^3  - Q_q \sin^2\theta_W  \right) q_L 
-  \bar q_R \gamma_\mu Q_q\sin^2 \theta_W q_R 
\right.  \\
&+
\left.  
\bar q^M_R \gamma_\mu \left( T^3  - Q_q \sin^2\theta_W \right) q^M_R 
-  \bar q^M_L \gamma_\mu Q_q\sin^2 \theta_W q_L^M 
\right] \; ,
\end{aligned}
\end{equation}
\begin{equation}
J^{\rm EM}_\mu =  Q_q \left( \bar q \gamma_\mu q 
+ \bar q^M \gamma_\mu q^M \right) \; .
\end{equation}

The above neutral current interactions in SM and the new Yukawa couplings
can induce FCNC decay $t \to Vq$ at one-loop level as 
depicted by the three Feynman diagrams in Fig.~\ref{FeynmanDiagram}.



\begin{figure}[!phtb]
\minipage{0.32\textwidth}
\begin{fmffile}{loop_A}
		\begin{fmfgraph*}(120,100)
			\fmfleft{i1}
			\fmfright{o1}
			\fmfbottom{b1,b2,b3,b4,b5}
			\fmftop{t1,t2,t3,t4,t5}
			\fmf{fermion,label=$t$}{i1,v1}
			\fmf{fermion}{v1,v2}
			\fmf{fermion,label=$q_m^M$}{v2,v3}
			\fmf{fermion,label=$q$}{v3,o1}
			\fmf{dashes,label=$\phi_{kS}$,left=1,tension=0.1}{v2,v3} 
			\fmf{wiggly,label=$Z,,\gamma$}{v1,b2}
			\fmf{phantom}{t2,v1}
		\end{fmfgraph*}
		\end{fmffile}
  \caption*{\small (A)}
\endminipage\hfill
\minipage{0.32\textwidth}
		\begin{fmffile}{loop_B}
		\begin{fmfgraph*}(120,100)
			\fmfleft{i1}
			\fmfright{o1}
			\fmfbottom{b1,b2,b3,b4,b5}
			\fmftop{t1,t2,t3,t4,t5}
			\fmf{fermion,label=$t$}{i1,v1}
			\fmf{fermion,label=$q^M_m$}{v1,v2}
			\fmf{fermion}{v2,v3}
			\fmf{fermion,label=$q$}{v3,o1}
			\fmf{dashes,label=$\phi_{kS}$,left=1,tension=0.1}{v1,v2} 
			\fmf{wiggly,label=$Z,,\gamma$}{v3,b4}
			\fmf{phantom}{t4,v3}
		\end{fmfgraph*}
		\end{fmffile}
  \caption*{\small (B)}
  \endminipage\hfill
\minipage{0.32\textwidth}%
      \begin{fmffile}{loop_C}
		\begin{fmfgraph*}(150,100)
			\fmfleft{i1}
			\fmfright{o1}
			\fmfbottom{b}
			\fmftop{t}
			\fmf{fermion,label=$t$}{i1,v1}
			\fmf{fermion,label=$q^M_m$}{v1,v2}
			\fmf{fermion,label=$q^M_m$}{v2,v3}
			\fmf{fermion,label=$q$}{v3,o1}
			\fmf{dashes,label=$\phi_{kS}$,left=1,tension=0.2}{v1,v3} 
			\fmf{wiggly,label=$Z,,\gamma$}{v2,b}
			\fmf{phantom}{t,v2}
		\end{fmfgraph*}
		\end{fmffile}
  \caption*{\small (C)}
  \endminipage
\caption{\small Feynman diagrams contributing to $t \to Vq$.}
\label{FeynmanDiagram}
\end{figure}

\section{FCNC Top Decays $t \to Vq$}
\label{calculation}

The effective Lagrangian for $t \to Z q$ and $t \to \gamma q$ can be expressed as
\begin{eqnarray}
\label{Leff}
{\mathcal L}_{\rm eff} & = & - \bar q \gamma_\mu ( C_L P_L + C_R P_R ) t Z^\mu 
- \frac{1}{m_t} \bar q \sigma_{\mu \nu} ( A_L P_L + A_R P_R ) t Z^{\mu\nu} \nonumber \\
& & \;\;\;\;\;\;\; -\frac{1}{m_t} \bar q \sigma_{\mu \nu} ( A^\prime_L P_L + A^\prime_R P_R ) t F^{\mu\nu} 
+ {\rm H.c.}
\end{eqnarray}
where $q=(u,c)$;
$Z^{\mu\nu} = \partial^\mu Z^\nu - \partial^\nu Z^\mu$ and 
$F^{\mu\nu} = \partial^\mu A^\nu - \partial^\nu A^\mu$;  
and $A_{L,R}$, $A^\prime_{L,R}$ and $C_{L,R}$ are dimensionless quantities.

In terms of the dimensionless mass ratios 
\begin{equation}
r_q \equiv m_q/m_t \; \; ,  \quad \quad r_Z \equiv m_Z/m_t 
\end{equation}
the partial decay rate for $t \to Zq$ is given by 
\begin{equation}
\Gamma ( t \to q Z ) = \frac{1}{16 \pi} \frac{1}{m_t}
\lambda^{\frac{1}{2}}\left(1, r_q^2, r_Z^2\right)
\left\langle \sum \vert {\mathcal M} \vert^2 \right\rangle \; ,
\end{equation}
where
$\lambda(1,y,z)=(1- (\sqrt y + \sqrt z)^2)(1- (\sqrt y - \sqrt z)^2)$
and
\begin{eqnarray}
\left\langle \sum \vert {\mathcal M} \vert^2 \right\rangle &=& 
\frac{m_t^2}{2} \biggl\{ 
+2 \left( \vert C_L \vert^2 + \vert C_R \vert^2 \right) \left( 1 + r_q^2 - r_Z^2 \right) \biggr.\nonumber \\
&+&  4 \left( \vert A_L \vert^2 + \vert A_R \vert^2 \right) 
\left[  2 \left( 1 - r_q^2 \right)^2 - \left( 1 + r_q^2 \right) r_Z^2 - r_Z^4 \right] \nonumber \\
&-& 16 \, {\rm Re} \left(C_L C_R^* \right) r_q - 48 \, {\rm Re} \left(A_L A_R^* \right) r_q r_Z^2 \nonumber \\
&-&12 \, {\rm Re} \left( C_L A_L^* + C_R A_R^* \right) r_q \left( 1 - r_q^2 + r_Z^2 \right) \\
&+&12 \, {\rm Re} \left( C_L A_R^* + C_R A_L^* \right) \left( 1 - r_q^2 - r_Z^2 \right) \nonumber \\
&+&\frac{1}{r_Z^2} \biggl. \left[ 
\left( \vert C_L \vert^2 + \vert C_R \vert^2 \right) 
\left( \left( 1 - r_q^2 \right)^2 - \left(1 + r_q^2 \right) r_Z^2 \right) 
+ 4  \, {\rm Re} \left( C_L C_R^* \right) r_q r_Z^2
\right] \biggr\}\, .\nonumber
\end{eqnarray}

For a given model, the dimensionless quantities $A_{L,R}$ and $C_{L,R}$ can be determined. 
In the mirror fermion model, these quantities are induced at one loop level, as depicted by the 
Feynman diagrams in Fig.~\ref{FeynmanDiagram}. 
Their analytical expressions are given in the Appendix.

Similarly, for  $t \to \gamma q$ we have
\begin{equation}
\Gamma ( t \to q \gamma ) = \frac{1}{16 \pi} \frac{1}{m_t} 
\lambda^{\frac{1}{2}}\left(1, r_q^2, 0\right)
\left\langle \sum \vert {\mathcal M} \vert^2 \right\rangle
\end{equation}
with $\left\langle \sum \vert {\mathcal M} \vert^2 \right\rangle
=4m_t^2(1-r_q^2)^2( \vert A^\prime_L \vert^2 + \vert A^\prime_R \vert^2)$. 
The expressions for $A^\prime_L$ and $A^\prime_R$ are also given in the Appendix.


\section{Analysis}
\label{analysis}

In our numerical analysis, we will make the following assumptions on the parameter space of the model.
\begin{itemize}

\item[(1)]
First, we will take all the unknown Yukawa couplings to be real and assume
$g_{iS}^q=g_{iS}^Q$ for $q=(u,d)$ and $i=0,1,2$. 
We will explore how our results depend on the couplings $g_{iS}^Q$.
We note that it has been shown recently~\cite{Hung:2017pss} in the extended mirror 
fermion model~\cite{Hoang:2014pda} the complex values of some of these Yukawa couplings, 
combined with the electroweak scale seesaw mechanism generating the minuscule neutrino masses, 
one can provide a solution to the strong CP problem without introducing axion.

\item[(2)]
Since only the product $V_{\rm CKM}=(V^u_L)^\dagger V_L^d$ are known experimentally, 
we will study the following scenarios for illustrative purpose.

Scenario 1:
\begin{equation}
\begin{aligned}
V_L^u &= V_{\rm CKM}^\dagger \; , \\
V_R^u = V_L^{u^M} = V_R^{u^M}  & = 1 \; .
\end{aligned}
\end{equation}

Scenario 2:
\begin{equation}
\begin{aligned}
V_L^u = V_L^{u^M}  & = V_{\rm CKM}^\dagger \; , \\
V_R^u = V_R^{u^M} & = 1 \; .
\end{aligned}
\end{equation}

\item[(3)]
For the three generation of mirror quark masses, we assume 
\begin{equation}
m_{q^M_1} : m_{q^M_2} : m_{q^M_3} = M : M + 10 \; {\rm GeV} : M+ 20 \; {\rm GeV} \; , 
\end{equation}
and vary the common mirror quark mass $M$ from 150 to 800 GeV.  
We note that mirror fermions in this class of electroweak scale mirror fermion model 
are expected to have masses of electroweak scale to satisfy unitarity~\cite{Chakdar:2015sra}.
 
\item[(4)]
For the scalars $\phi_{kS}$, their masses are necessarily small since they are link 
to the Dirac neutrino masses~\cite{Hung:2015nva,Hung:2006ap}. 
We set their masses $m_{kS}$ all equal 10 MeV.  Our numerical results 
are not sensitive to this choice as long as $m_{kS} \ll m_{q^M_m}$.

\item[(5)]
For the SM parameters, we use~\cite{PDG}
\begin{eqnarray}
m_t & = &173.21\, {\rm GeV} \; , \; m_c = 1.275 \, {\rm GeV} \; , \;  m_u = 2.3 \, {\rm MeV} \; ,  \nonumber\\
\sin^2 \theta_W & = & 0.23126 \; , \; \alpha = 1/127.944 \; ,  \\
\Gamma_t & = & 1.41\, {\rm GeV} \; , \; {\cal B}\left( t \to W^+b \right) = 0.957 \; . \nonumber
\end{eqnarray}
\end{itemize}


\begin{figure}[!phtb]
\minipage{0.45\textwidth}
\includegraphics[width=\linewidth]{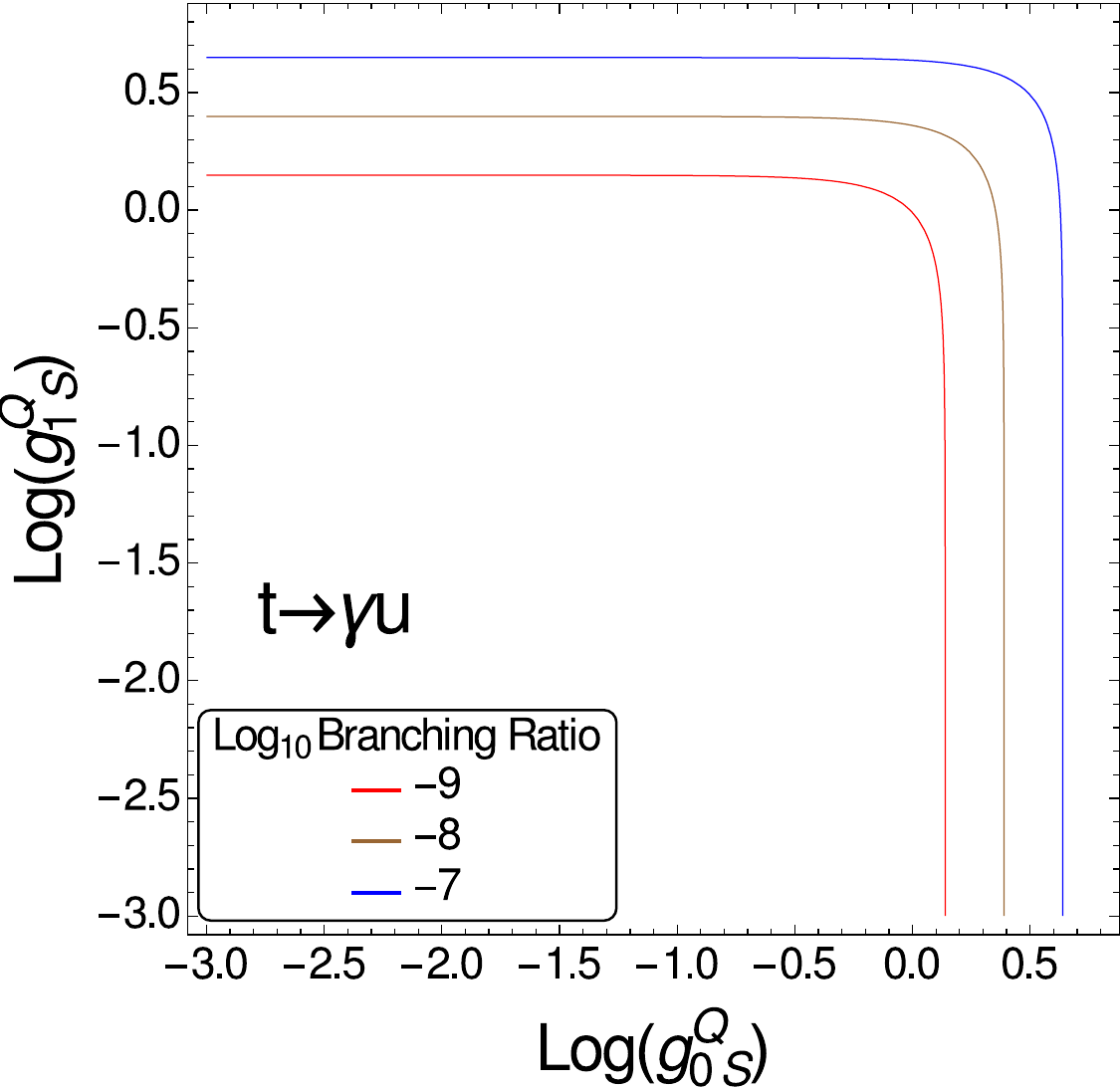}
\endminipage\hfill
\minipage{0.45\textwidth}
  \includegraphics[width=\linewidth]{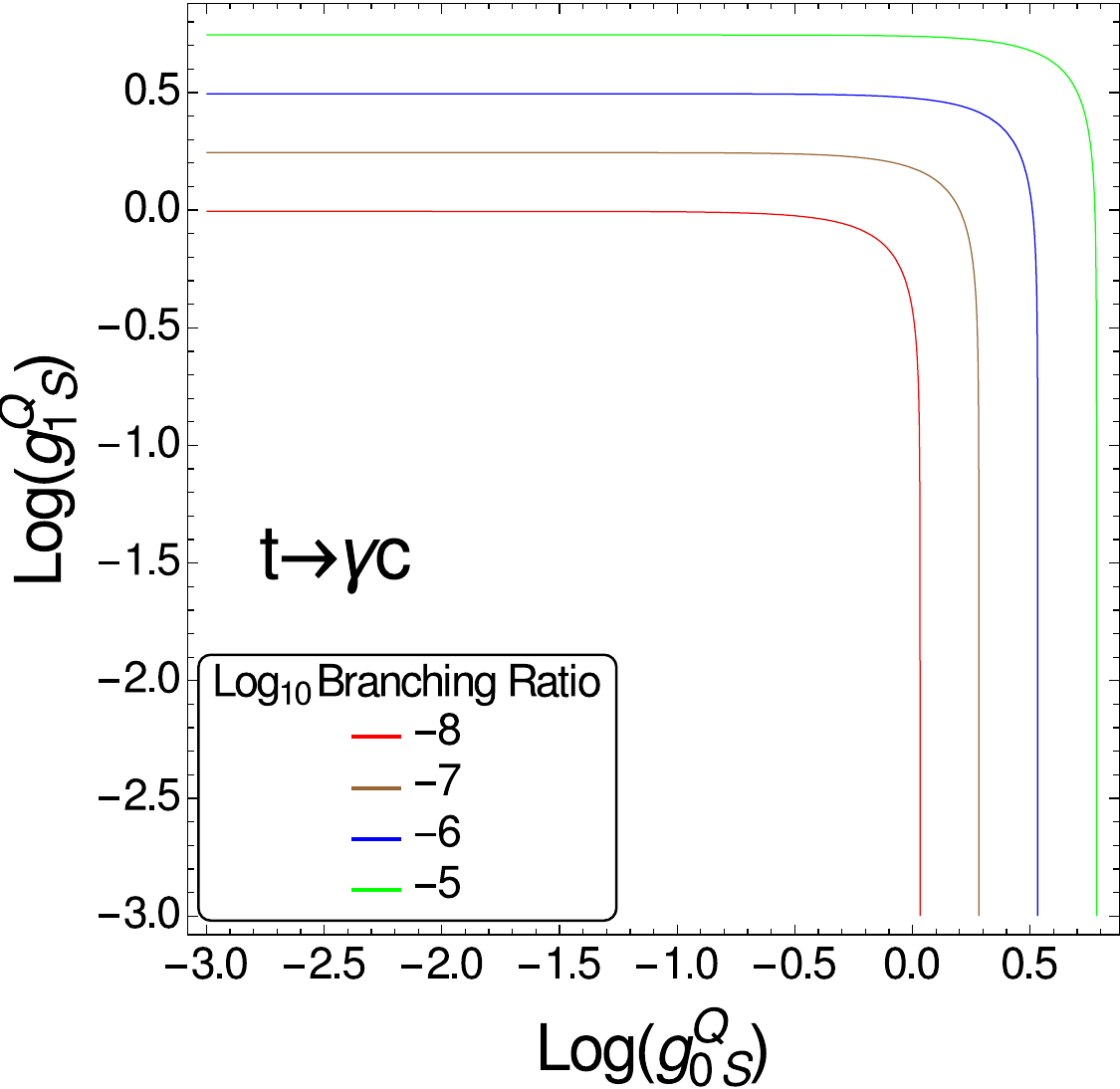}
  \endminipage\hfill
\caption{\small Branching ratios of $t \to \gamma u$ (left) and $t \to \gamma c$ (right)
versus the logarithmic of Yukawa couplings $g^Q_{0S}$ and $g^Q_{1S}$ with 
$g^Q_{2S} = 10^{-3}$ and $M = 150$ GeV in Scenario 1.}
\label{BRtGq_scenario1}
\end{figure}

\begin{figure}[!phtb]
\minipage{0.45\textwidth}
  \includegraphics[width=\linewidth]{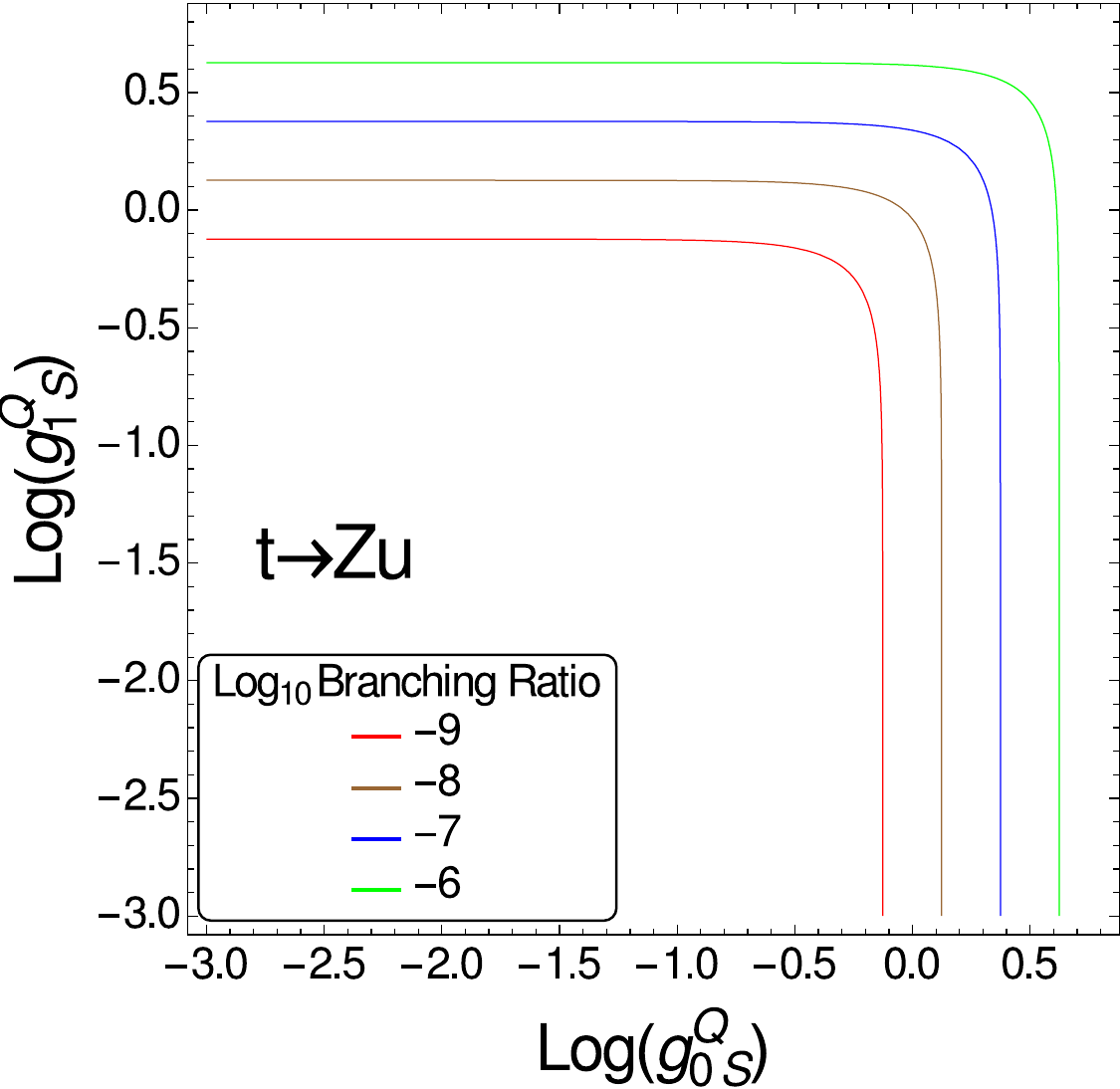}
 \endminipage\hfill
\minipage{0.45\textwidth}
  \includegraphics[width=\linewidth]{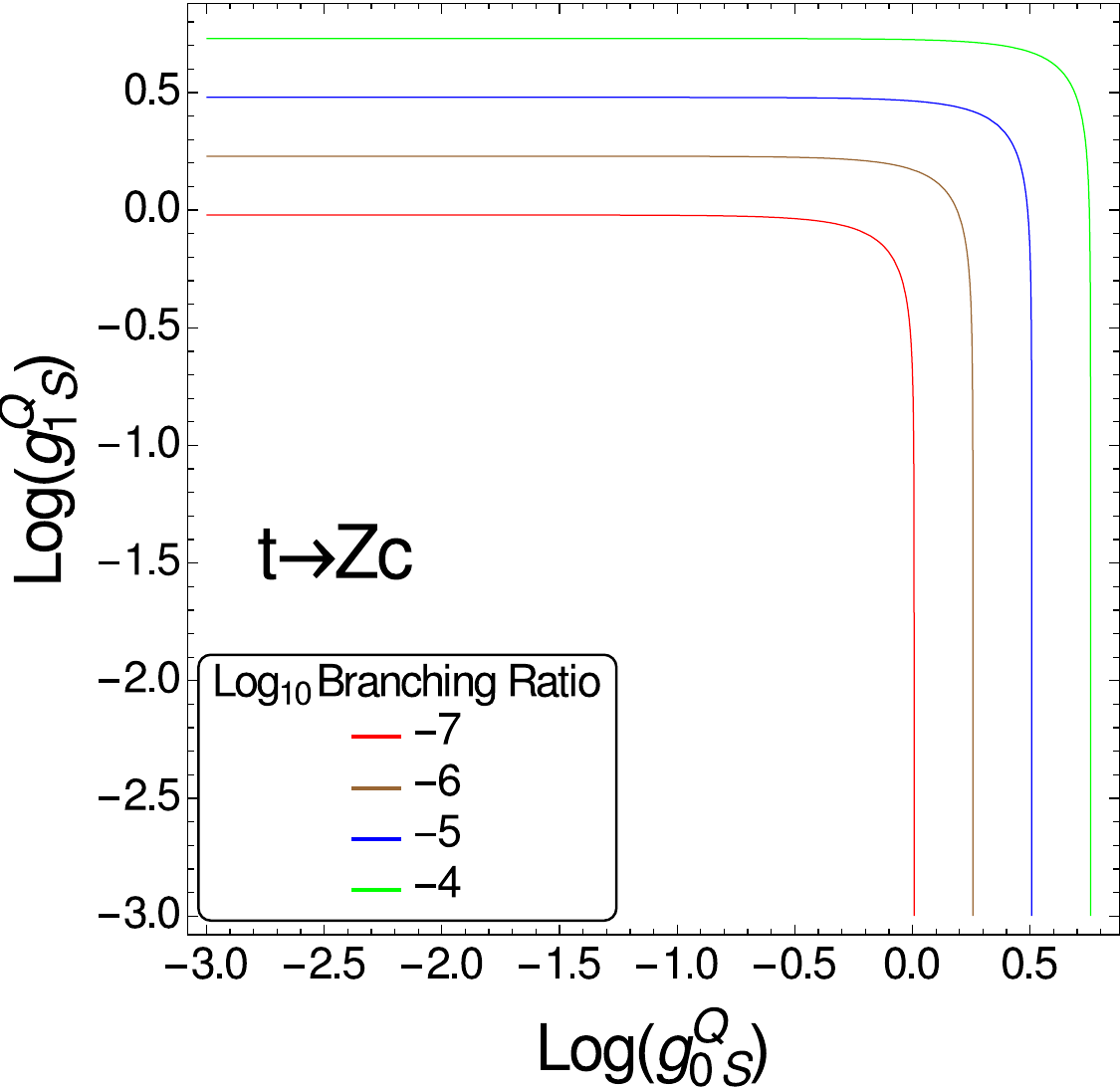}
    \endminipage\hfill
\caption{\small Branching ratios of $t \to Z u$ (left) and $t \to Z c$ (right)
versus the logarithmic of Yukawa couplings $g^Q_{0S}$ and $g^Q_{1S}$ with 
$g^Q_{2S} =10^{-3}$ and $M = 150$ GeV in Scenario 1.}
\label{BRtZq_scenario1}
\end{figure}

\begin{figure}[!phtb]
\minipage{0.45\textwidth}
  \includegraphics[width=\linewidth]{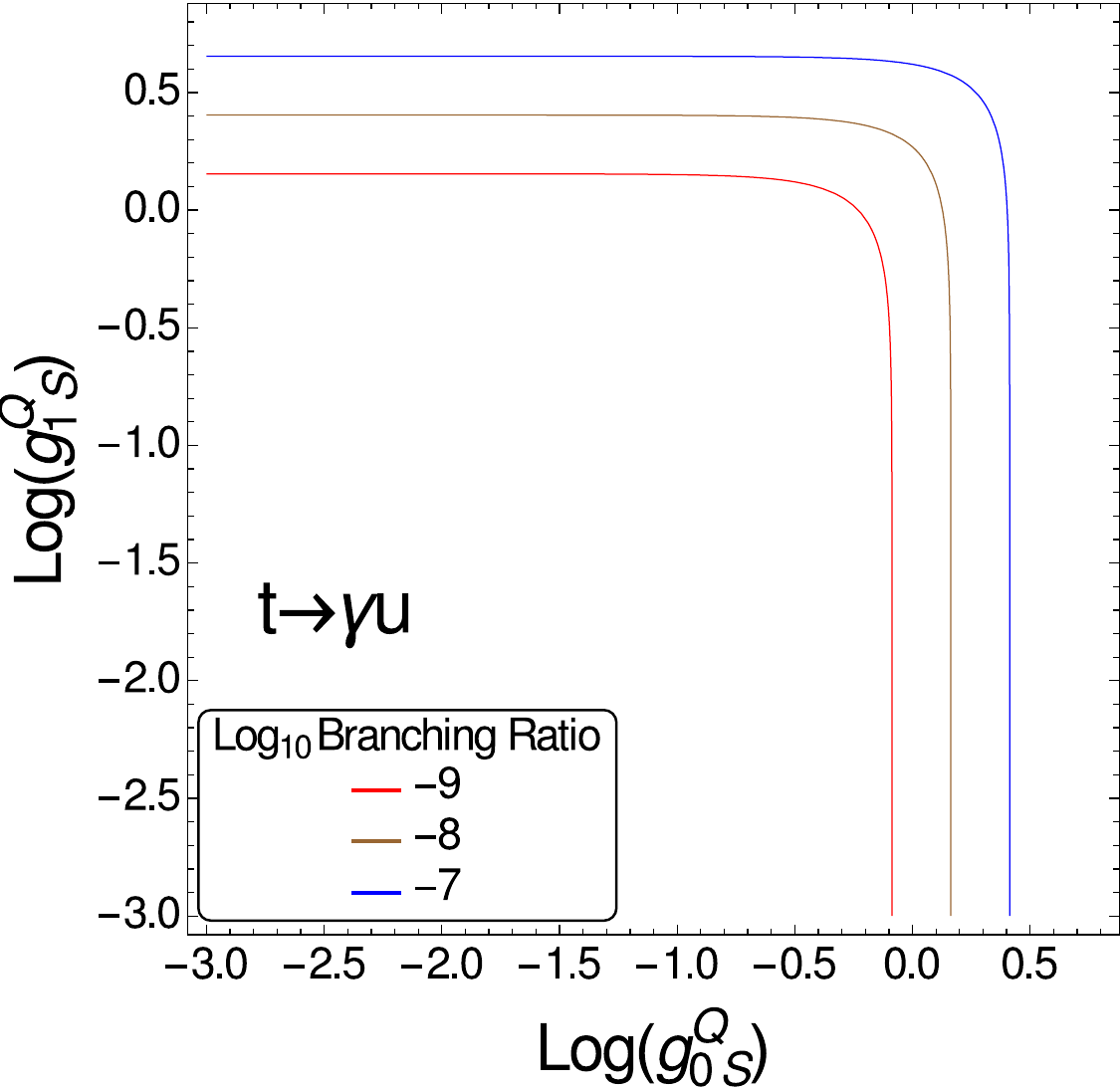}
\endminipage\hfill
\minipage{0.45\textwidth}
  \includegraphics[width=\linewidth]{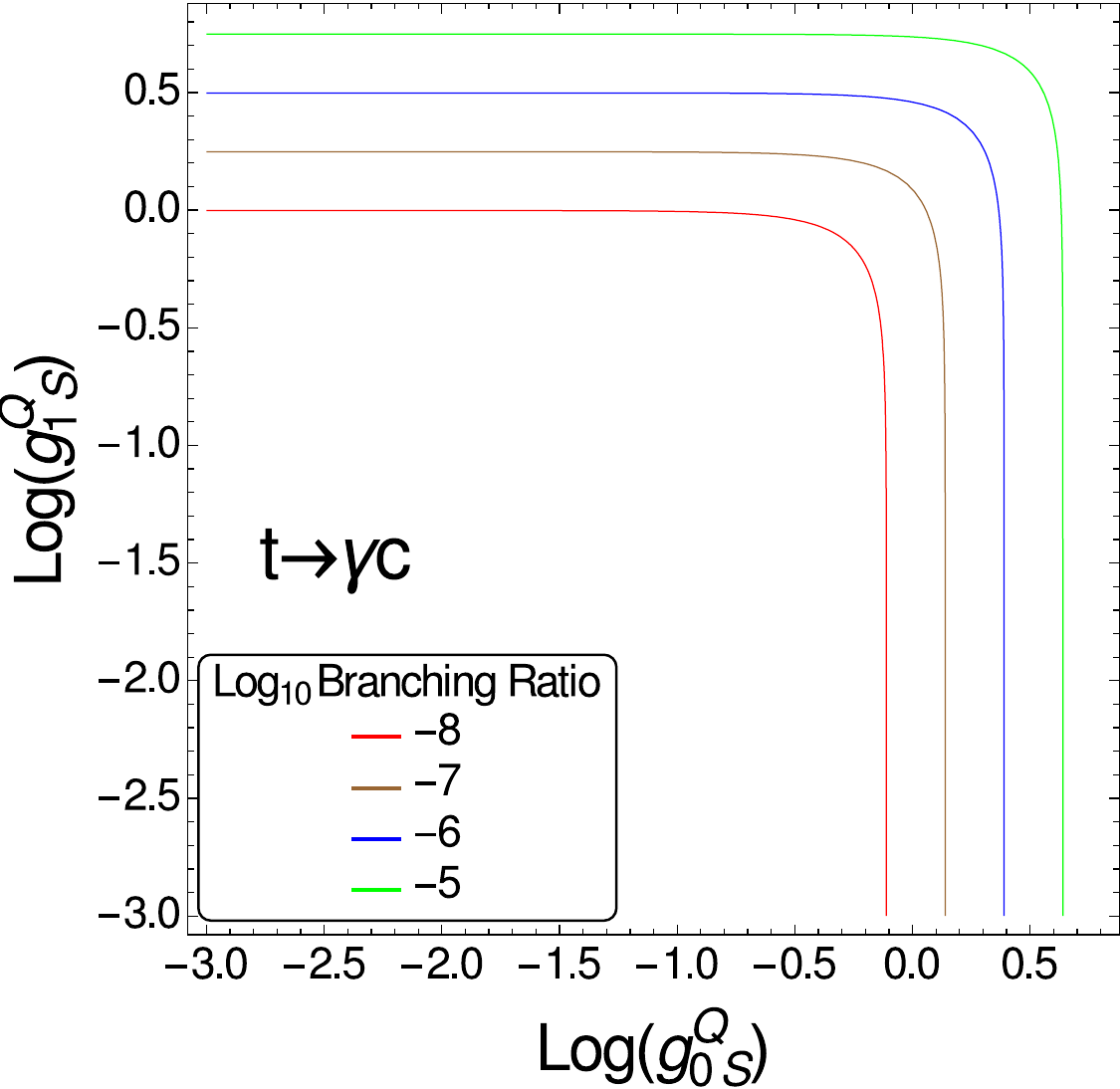}
    \endminipage\hfill
\caption{\small Same as Fig.~(\ref{BRtGq_scenario1}) in Scenario 2.}
\label{BRtGq_scenario2}
\end{figure}

\begin{figure}[!phtb]
\minipage{0.45\textwidth}
  \includegraphics[width=\linewidth]{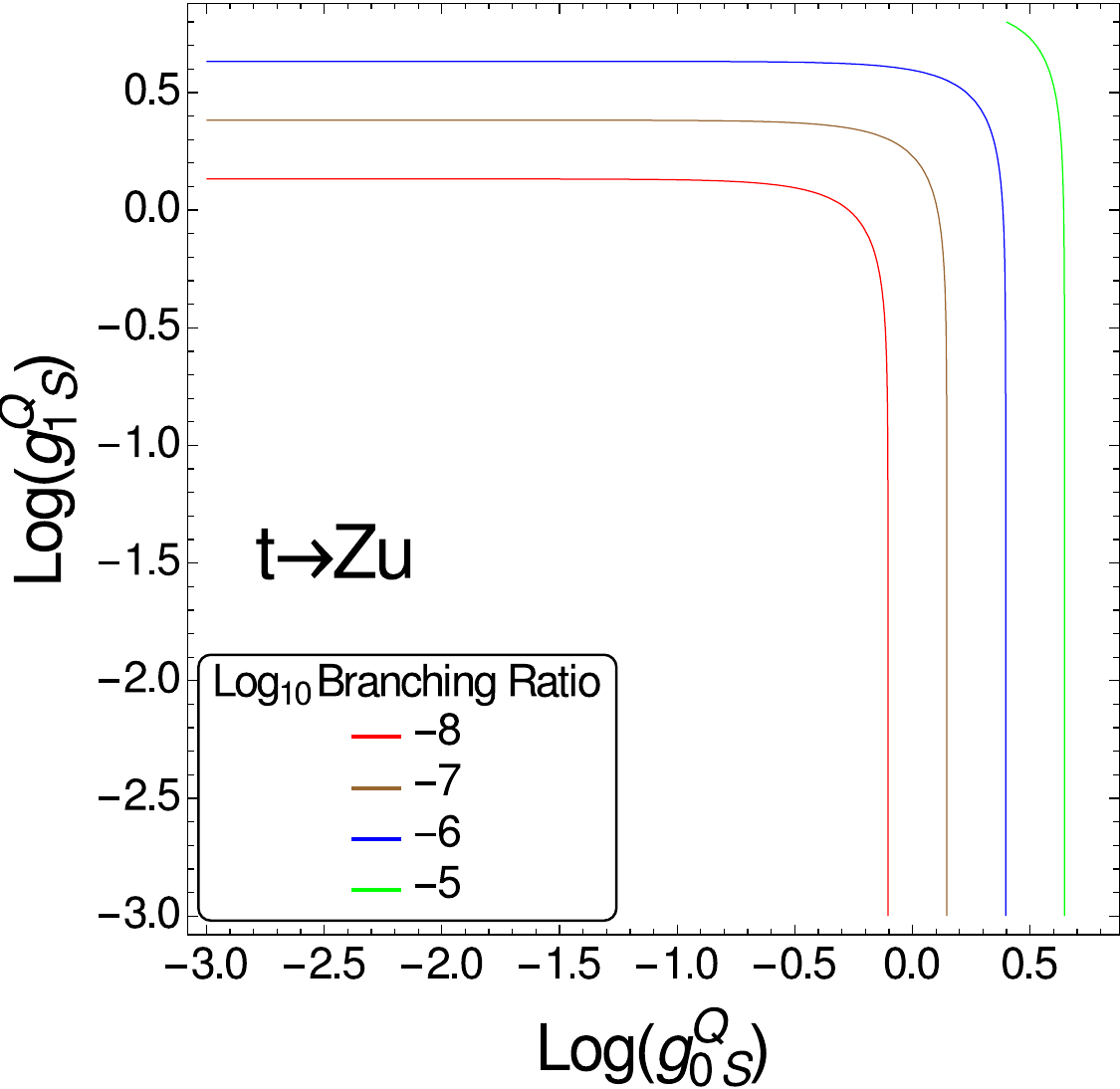}
\endminipage\hfill
\minipage{0.45\textwidth}
  \includegraphics[width=\linewidth]{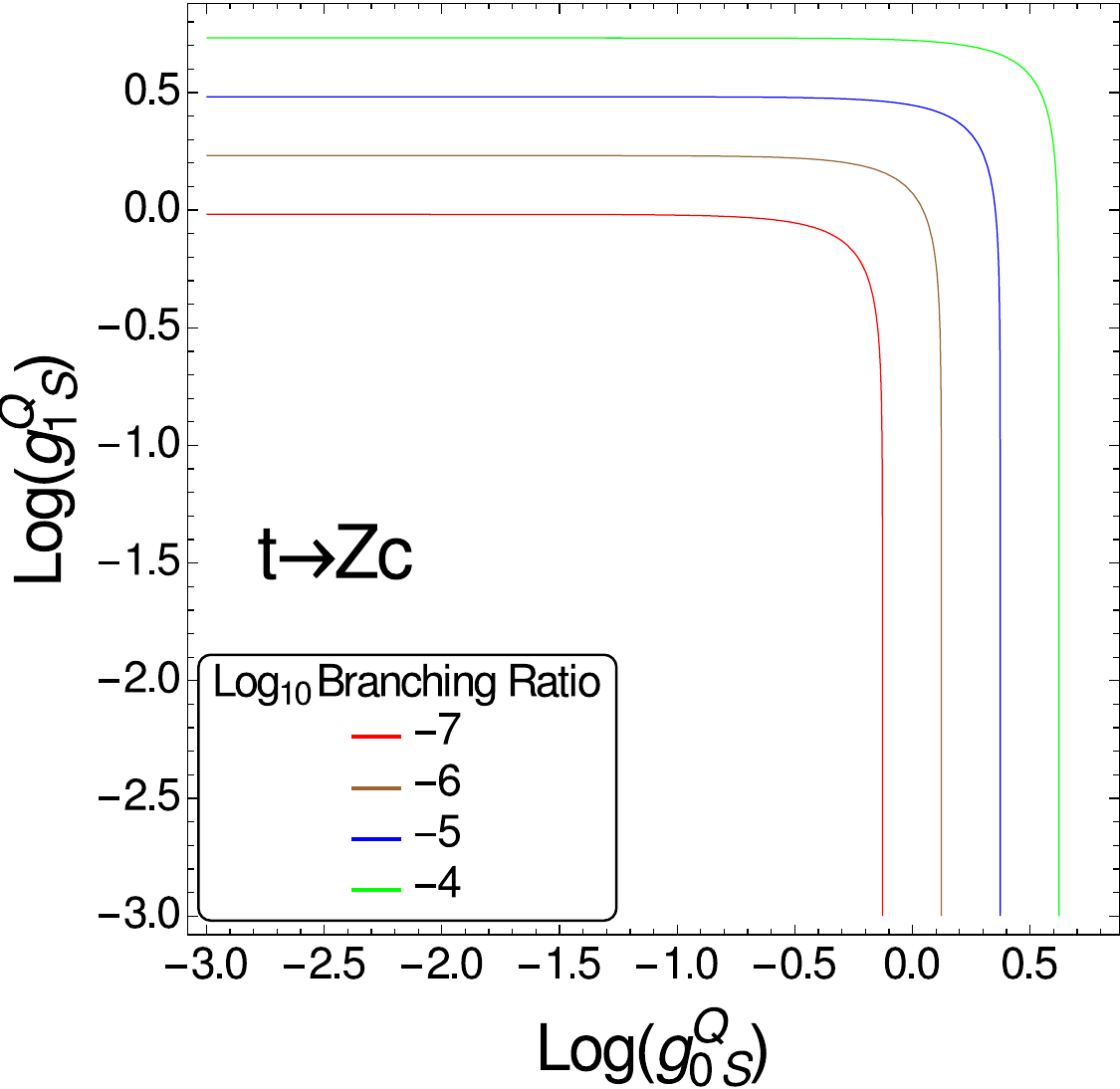}
  \endminipage\hfill
\caption{\small Same as Fig.~(\ref{BRtZq_scenario1}) in Scenario 2.}
\label{BRtZq_scenario2}
\end{figure}

In Fig.~(\ref{BRtGq_scenario1}), we show the contour plots of 
$\log {\cal B}(t \to \gamma u)$ (left panel) and 
$\log {\cal B}(t \to \gamma c)$ (right panel) on the $(\log_{10} (g_{0S}^Q), \log_{10} (g_{1S}^Q))$ plane in the case of Scenario 1 with $g^Q_{2S}$ set to be $10^{-3}$.
Fig.~(\ref{BRtZq_scenario1}) is similar as Fig.~(\ref{BRtGq_scenario1}) but for 
$t \to Zq$.
Figs.~(\ref{BRtGq_scenario2}) and (\ref{BRtZq_scenario2}) are the same as Figs.~(\ref{BRtGq_scenario1}) and (\ref{BRtZq_scenario1}) respectively 
but for Scenario 2. The common mirror fermion mass $M$ is 
set to be 150 GeV in these 4 figures.

\begin{figure}[!phtb]
\minipage{0.45\textwidth}
  \includegraphics[width=\linewidth]{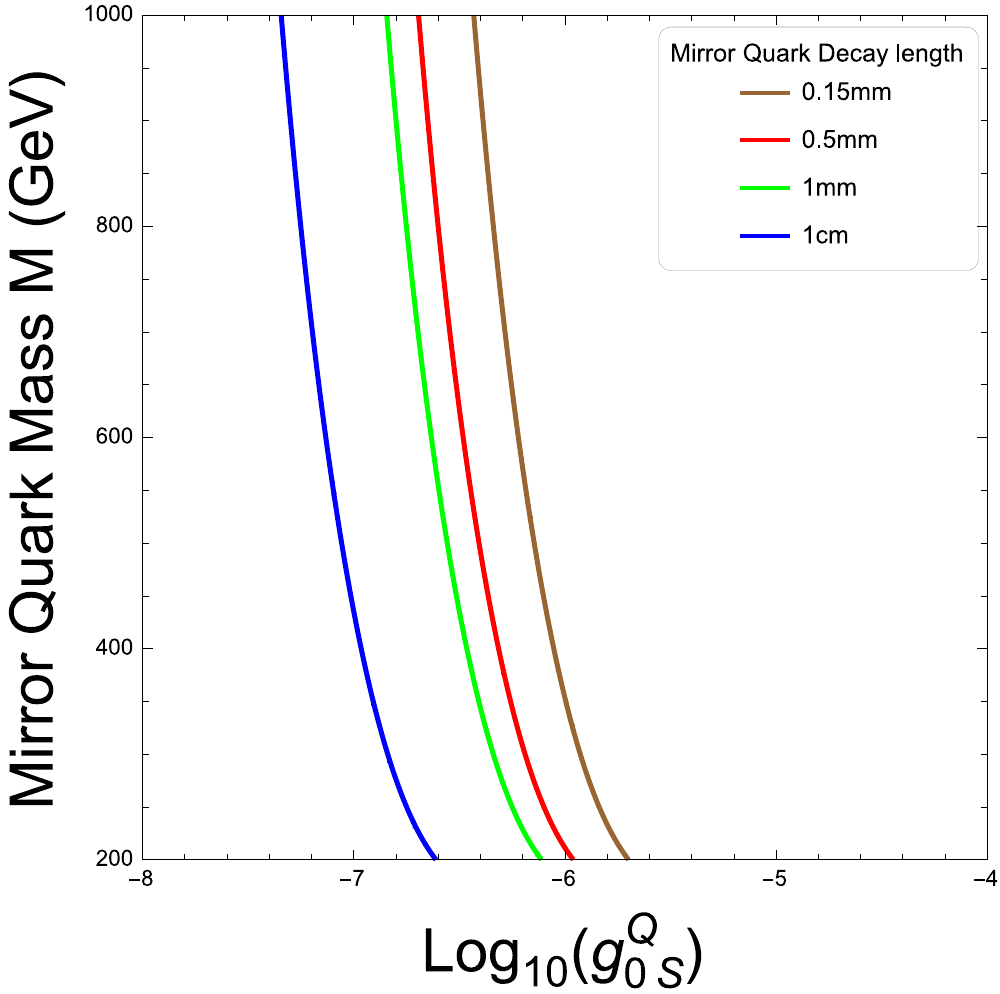}
\endminipage
\caption{\small Decay length of the lightest mirror quark versus 
$\log_{10} (g^Q_{0S})$ assuming all the unknown Yukawa couplings equal to each other.
}
\label{Mtopdecaylength}
\end{figure}

Mirror quarks may be pair produced at the LHC~\cite{Chakdar:2015sra}. Once produced the heavier mirror fermions may cascade into lighter ones by emitting 
an on-shell or off-shell SM $W$-boson, depending on the detail mass spectrum of the mirror fermions. 
The lightest mirror quark will then decay into its SM partner with 
any one of the scalar singlets $\phi_{kS}$  via the new Yukawa interactions which are 
responsible to the FCNC decays of the top quark studied here. 
In the mirror lepton case, the corresponding Yukawa couplings $g_{iS}^L$ are necessarily small since they are responsible for providing small Dirac masses to the neutrinos in the electroweak scale seesaw mechanism. 
Assuming the lightest mirror fermion is $u^M$. In Fig.~(\ref{Mtopdecaylength}), we plot the
contours of the decay length of $u^M$ in the $(M, \log_{10} (g_{0S}^Q))$ plane. We take all the Yukawa couplings to be the same just for illustrations. One can see that for very small Yukawa couplings  $<10^{-6}$, can the decay length 
reach a few mm for a displaced vertex. 
Search strategies for the mirror fermions would then be quite different from the usual cases, involving not merely the 
missing energies but displaced vertices as well~\cite{Chakdar:2015sra}. 
Current experiments at the LHC have the capability to perform such kind of searches.
Further studies of this issue are warranted.

Nevertheless, for the mirror quarks, there is no {\it a priori} reason that these new Yukawa couplings have to be very small except that there are stringent 
constraints from the mixings between SM fermions and their mirrors. 
The mixing angle is roughly of order $g^Q_{iS} \langle \phi_{kS} \rangle / M$. 
For $g^Q_{iS} \sim 1$, $\langle \phi_{kS} \rangle \sim 1$ MeV and $M \sim$ 500 GeV, this mixing angle is about $2 \times 10^{-6}$.
A full analysis taking into the account of the mixing effects is beyond the scope of this paper.

\begin{figure}[!phtb]
\minipage{0.45\textwidth}
  \includegraphics[width=\linewidth]{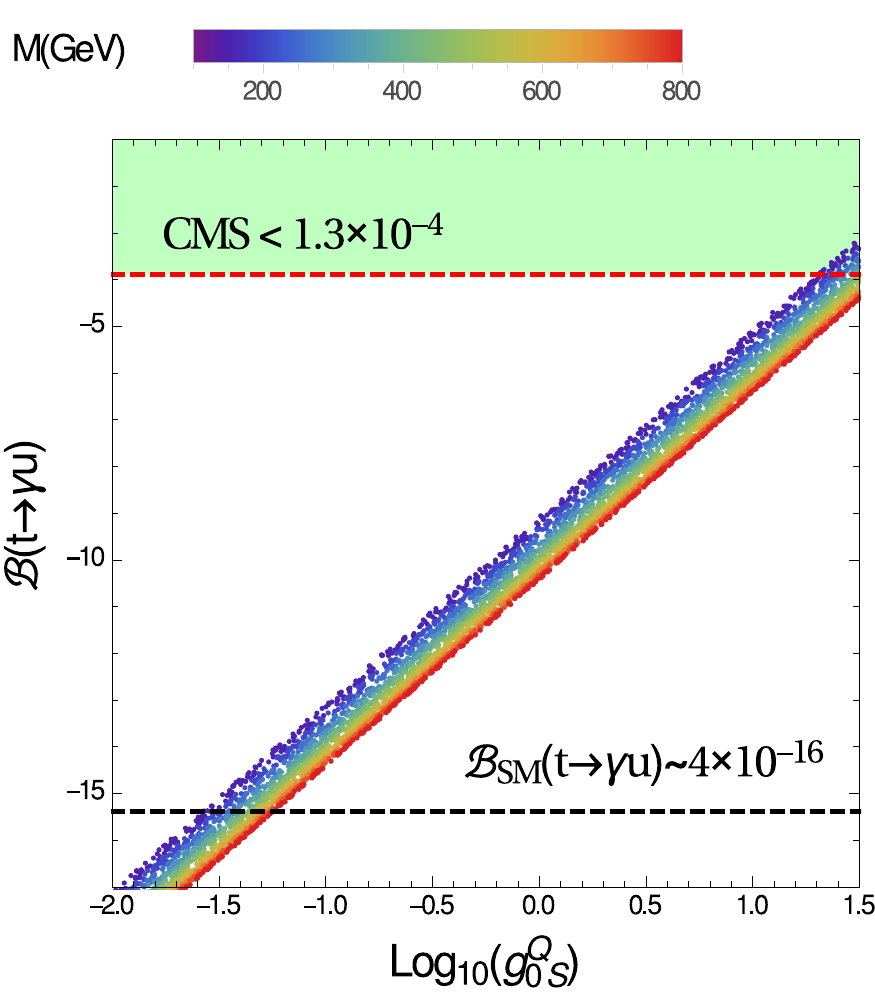}
  \includegraphics[width=\linewidth]{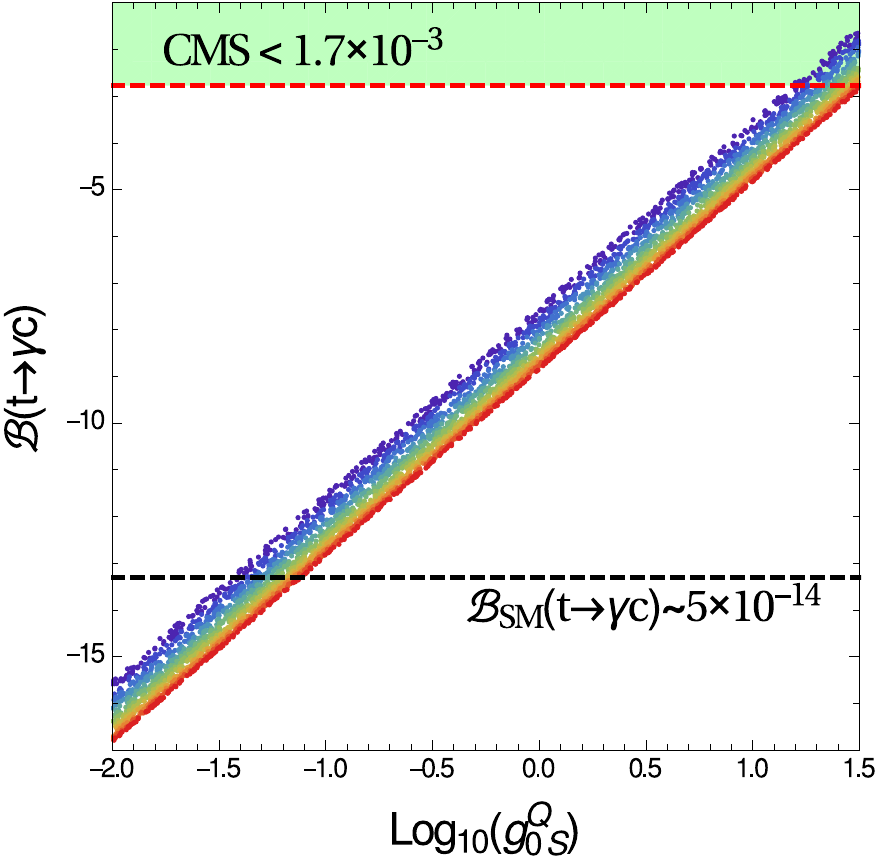}
\endminipage
\minipage{0.45\textwidth}
  \includegraphics[width=\linewidth]{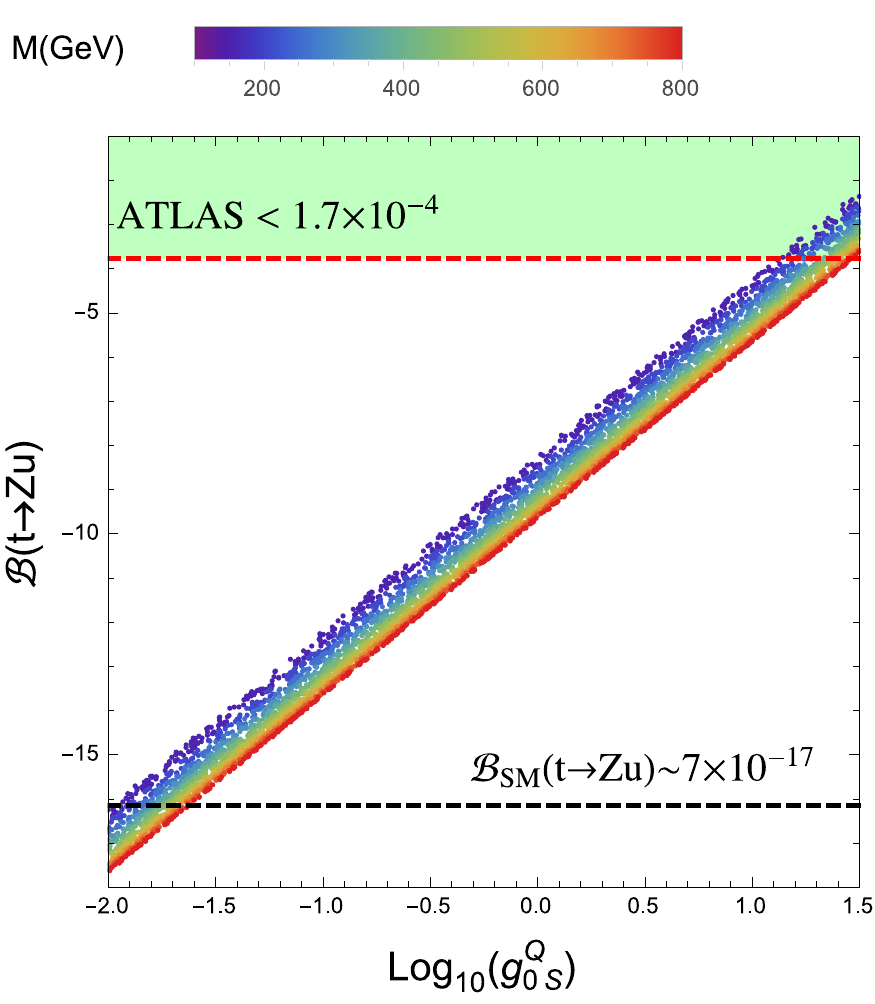}
  \includegraphics[width=\linewidth]{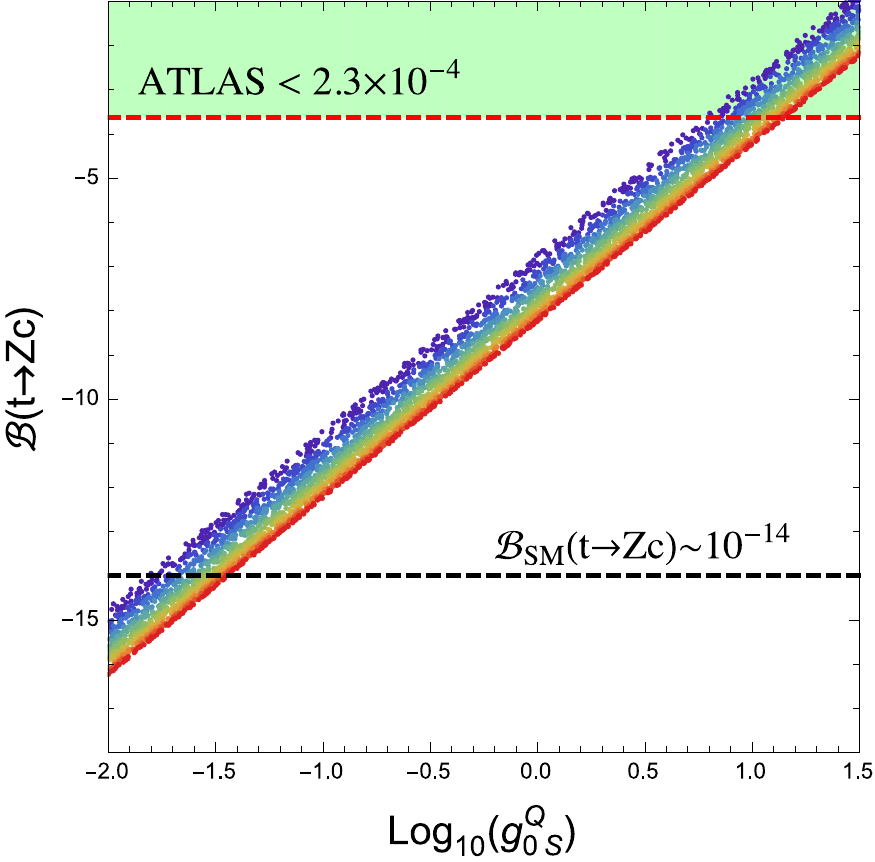}
\endminipage
  \caption{\small Scatter plots for the branching ratios of $t \to Vq$ in Scenario 1.}
  \label{Scatter plot scenario1}
\end{figure}

\begin{figure}[!phtb]
\minipage{0.45\textwidth}
  \includegraphics[width=\linewidth]{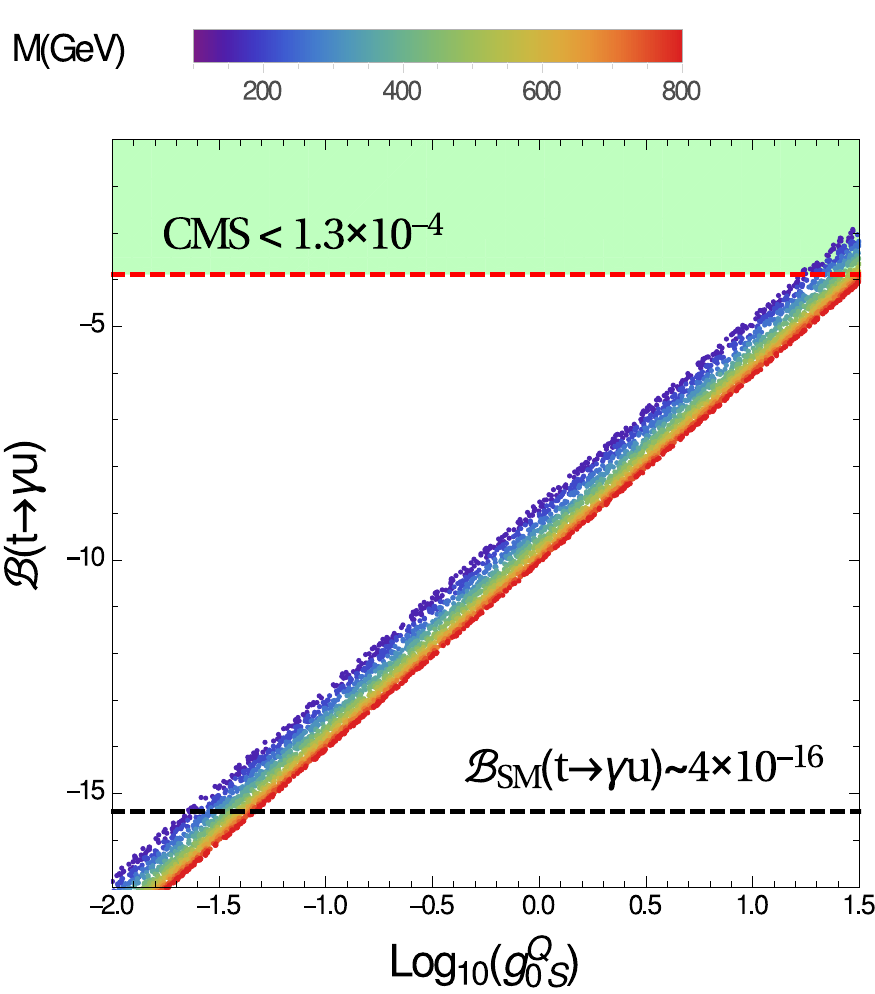}
  \includegraphics[width=\linewidth]{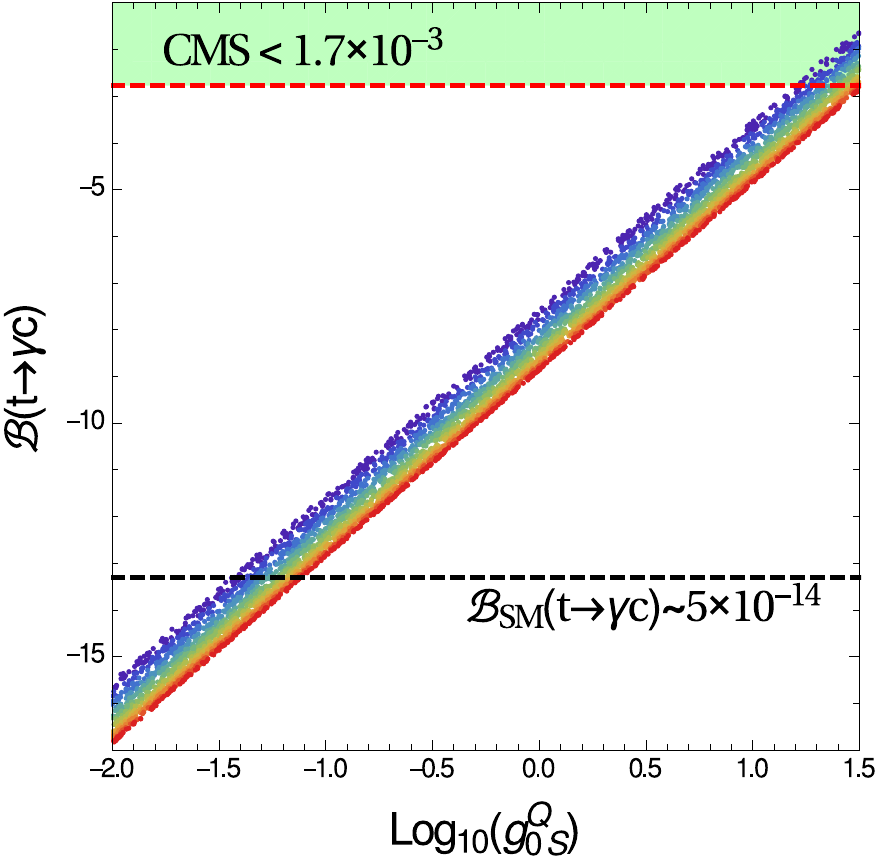}
\endminipage
\minipage{0.45\textwidth}
  \includegraphics[width=\linewidth]{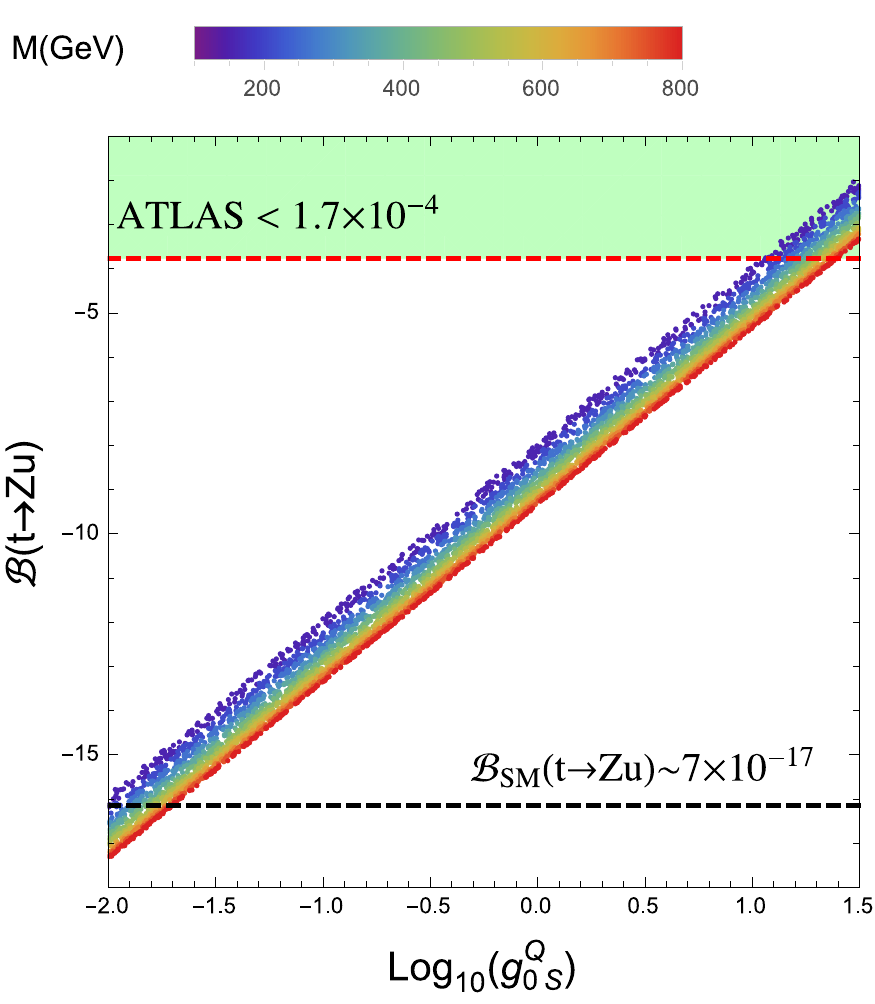}
  \includegraphics[width=\linewidth]{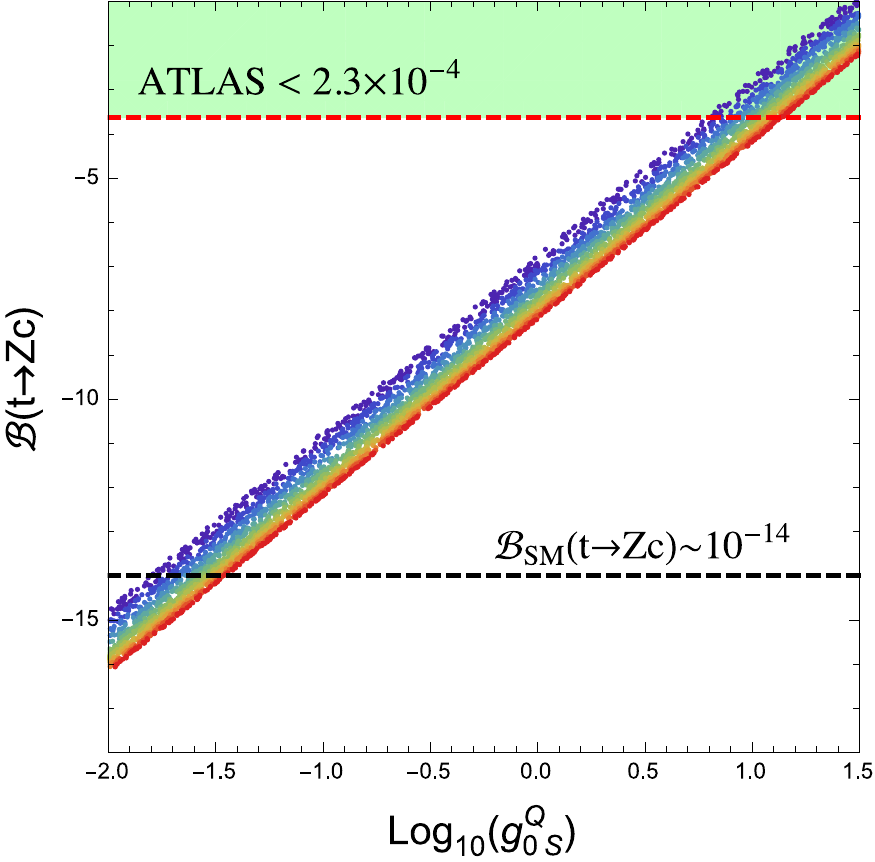}
\endminipage
  \caption{\small Same as Fig.~(\ref{Scatter plot scenario1}) in Scenario 2.}
  \label{Scatter plot scenario2}
\end{figure}

In Figs.~(\ref{Scatter plot scenario1}) and (\ref{Scatter plot scenario2}), we show the scatter plots for the logarithmic of branching ratios of the 4 processes, 
$t \to \gamma u$ and $t \to \gamma c$ in the left panel and 
$t \to Z u$ and $t \to Z c$ in the right panel, versus $\log_{10} (g_{0S}^Q)$ 
for Scenarios 1 and 2 respectively.
We have set all the Yukawa couplings equal to each other 
in these plots.
The different colors in the scatter plots represent different values of the common 
mirror fermion mass $M$ varied 
from 150 to 800 GeV as indicated by the color palettes 
on the top of each plot. 
Current experimental limits of these processes are also shown in these plots by the horizontal red dashed lines, while the black dashed lines are the SM predictions.
It is clear from these plots that the mirror quarks in this class of model with mass less than 
800 GeV could play an important role in FCNC decays of the top quark, 
provided that the Yukawa couplings are of the same size as the top quark Yukawa coupling in SM.
However if the Yukawa couplings are very small to allow for
a displaced vertex for the lightest mirror fermion, all these FCNC top decays are beyond the reach at LHC.

\begin{table}[h!]
\begin{tabular}{|c|c|c|c|c|c|}
\hline
 Process&SM&2HDM&MSSM&Extra-Dimension&Mirror Model\\
 \hline
 $t{\rightarrow}Zc$&$1\times10^{-14}$&$\le10^{-10}(10^{-6})$&$\le10^{-7}(10^{-6})$&$\le10^{-5}$&${10^{-6} - 10^{-8}}$\\
 $t{\rightarrow}Zu$&$7\times10^{-17}$&$-$&$\le10^{-7}(10^{-6})$&$-$&${10^{-8} - 10^{-10}}$\\
 $t{\rightarrow}{\gamma}c$&$5\times10^{-14}$&$\le10^{-9}(10^{-7})$&$\le10^{-8}(10^{-9})$&$\le10^{-9}$&${10^{-7} - 10^{-9}}$\\
 $t{\rightarrow}{\gamma}u$&$4\times10^{-16}$&$-$&$\le10^{-8}(10^{-9})$&$-$&${10^{-8} - 10^{-11}}$\\
 
\hline
\end{tabular}
\caption{\small 
Comparisons of theoretical predictions for the 
branching ratios of FCNC rare top decay $t \to Vq$ in various models. 
The numbers in brackets in 2HDM and MSSM are for 2HDM with tree level flavor violations and 
MSSM with $R$-parity violation respectively.}
\label{summaryBRs}
\end{table}

\section{Conclusion}
\label{summary}

In Table~\ref{summaryBRs}, we summarize our numerical results as well as those from SM and 
other three popular new physics models taken from~\cite{Agashe:2013hma} for comparisons.
The numbers in brackets in the 2HDM and MSSM columns 
are for 2HDM with tree level flavor violation and MSSM with $R$-parity violation respectively.
Our results shown in the last column are taken from Figs.~(\ref{Scatter plot scenario1}) and (\ref{Scatter plot scenario2}) 
for $g_{0S}^Q=y_t =\sqrt 2 m_t/v_{\rm SM} \sim 1$ and the mirror mass $M$ varying 
from 150 to 800 GeV.
On the other hand, if $\vert g_{0S}^Q \vert$ turns out to be small, of order $10^{-4}$ or less
as suggested by the new solution to the strong CP problem discussed in~\cite{Hung:2017pss},
all FCNC top decays in the model would be unobservable.

While the experimental results of many FCNC processes in the kaon, $D$ and $B$ meson systems, accumulated over the past several decades, had been mostly consistent 
with the SM expectations, theoretical predictions for 
FCNC processes involving the top quark and/or the Higgs boson have not been challenged by high energy experiments until recent years. 

In this work, we have computed the FCNC processes $t \to Zq$ and $\gamma q$ in a class of mirror fermion model equipped with a horizontal $A_4$ symmetry in the fermion and scalar sectors. We found that branching ratio for $t \to Zc$ is typically of order 
$10^{-8} - 10^{-6}$ as mirror quark masses are varying in the range 
from 800 to 150 GeV and new Yukawa couplings are of the 
same size of the top Yukawa coupling in SM.
At 14 TeV the total cross section for $t \bar t$ production at the LHC is about 598 pb. 
With a luminosity of 300 (1000) fb$^{-1}$, we thus expect 180 (6) events of $t \to Zc$ for a branching ratio of $10^{-6}$ ($10^{-8}$) before any experimental cuts.
For the other processes $t \to Zu$ and $t \to \gamma q$, their branching ratios are typically smaller by  
$1-2$ orders of magnitude. 

For the gluon mode $t \to g q$ its partial width is about 42 times larger than that of the photon mode $t \to \gamma q$.
The LHC limits for the branching ratios of $t \to g u$ and $t \to g c$ are  $2.0 \times 10^{-5}$ and $4.1 \times 10^{-4}$ respectively
 from CMS~\cite{Khachatryan:2016sib}, and 
 $4.0 \times 10^{-5}$ and $20 \times 10^{-5}$ respectively from ATLAS~\cite{Aad:2015gea}. 
These branching ratios are extracted from the single top production via FCNC interactions from 
gluon plus up- or charm-quark initial states. They are still $1-2$ orders of magnitude above our theoretical predictions.

It is also interesting to consider FCNC processes involving both the heavy top quark and the 125 GeV Higgs, the two heaviest particles in the SM. One particular important process is $t \to h q$, which LHC has obtained the following limits~\cite{Aad:2015pja,Khachatryan:2016atv}
\begin{equation}
{\mathcal B}(t \to hu)  \leq
\left\{ 
\begin{array}{l}
4.5 \times 10^{-3} \;  [{\rm ATLAS}] \; , \\
5.5 \times 10^{-3} \;  [{\rm CMS}] \; ; 
\end{array}
\right.
\end{equation}
\begin{equation}
{\mathcal B}(t \to hc)  \leq
\left\{
\begin{array}{l}
4.6 \times 10^{-3}  [{\rm ATLAS}] \; ,\\
4.0\times 10^{-3}  [{\rm CMS}] \; .
\end{array}
\right.
\end{equation}
In the mirror fermion model, the mirror Higgs as well as the GM triplets 
could be an imposter for the 125 GeV Higgs due to mixing effects,
which must be taken into account. This work will be reported elsewhere~\cite{wip}.

LHC has unique opportunity for probing the top quark FCNC decays in new physics models since the SM contributions are at least nine orders of magnitude below the current limits of these processes. With its high luminosity upgrade in the second phase, HL-LHC can impose powerful constraints on any underlying new physics responsible for the FCNC interactions. 
 
\section*{Acknowledgments}
We would like to thank Chuan-Ren Chen for stimulating discussions. 
This work is supported by the Ministry of Science and Technology (MoST) of Taiwan under
Grant Number MOST-104-2112-M-001-001-MY3. 

\newpage

\section*{Appendix}

\subsection*{Form Factors for $t \to Zq$}

All three Feynman diagrams (A), (B) and (C) in Fig.~(\ref{FeynmanDiagram}) contribute to the form factors $C_L$ and $C_R$: 
$$
C_{(L,R)}=C_{(L,R)}^A+C_{(L,R)}^B+C_{(L,R)}^C \; .
$$ 
To minimize cluttering in our expressions given below, we define
$$a=\frac { g }{ 4\cos{ \theta  }_{ W } } 
\left(1-\frac { 8 }{ 3 } { \sin }^{ 2 }{ \theta  }_{ W }\right) \; , \;\;\  \;\;\; b=\frac { g }{ 4\cos{ \theta  }_{ W } } \; ,$$
and
$$V^{L,k}_{ij} = \left( V^{u,k}_{L} \right)_{ij} \; , \;\;\  \;\;\;  V^{R,k}_{ij} = \left( V^{u,k}_{R} \right)_{ij} \; ,$$
where $V^{u,k}_{(L,R)}$ are given by Eq.~(\ref{VqkLR}).
The individual contributions from each diagrams can be computed using the automated tools \texttt{FormCalc} and \texttt{LoopTools} in \texttt{FeynArts}~\cite{feynarts}. 
The results are listed as follows.
\begin{equation}
    \begin{aligned}
        { C }_{ L }^{ A }=&\frac { 1 }{ { ( { m }_{ t } }^{ 2 }-{ { m }_{ q } }^{ 2 } ) } \frac { ( a+b ) }{ 16{ \pi  }^{ 2 } } \sum_{ k=0 }^{ 3 }\sum _{ m=1 }^{ 3 }\biggl\{ { { m }_{ q } }^{ 2 } { V }^{ L,k }_{qm}{ { V }^{ L,k }_{tm} }^{ * } \bigl[ B_0({ { m }_{ q } }^{ 2 },{ { m }_{ {\phi}_{kS} } }^{ 2 },{ { m }_{ { q }_{ m }^{ M } } }^{ 2 })\bigr. \biggr.\\
        &+B_1({ { m }_{ q } }^{ 2 },{ { m }_{ {\phi}_{kS} } }^{ 2 },{ { m }_{ { q }_{ m }^{ M } } }^{ 2 }) \big]+{ m }_{ t }{ m }_{ q } { V }^{ R,k }_{qm}{ { V }^{ R,k }_{tm} }^{ * } \big[B_0({ { m }_{ q } }^{ 2 },{ { m }_{ {\phi}_{kS} } }^{ 2 },{ { m }_{ { q }_{ m }^{ M } } }^{ 2 })\\
        &+B_1({ { m }_{ q } }^{ 2 },{ { m }_{ {\phi}_{kS} } }^{ 2 },{ { m }_{ { q }_{ m }^{ M } } }^{ 2 })\big]+{ m }_{ q }{ m }_{ { q }_{ m }^{ M } } { V }^{ R,k }_{qm}{ { V }^{ L,k }_{tm} }^{ * } B_0({ { m }_{ q } }^{ 2 },{ { m }_{ {\phi}_{kS} } }^{ 2 },{ { m }_{ { q }_{ m }^{ M } } }^{ 2 })\\
        &+\biggl. { m }_{ t }{ m }_{ { q }_{ m }^{ M } } { V }^{ L,k }_{qm}{ { V }^{ R,k }_{tm} }^{ * } B_0({ { m }_{ q } }^{ 2 },{ { m }_{ {\phi}_{kS} } }^{ 2 },{ { m }_{ { q }_{ m }^{ M } } }^{ 2 }) \biggr\}  \; ,
        \end{aligned}
\end{equation}
\begin{equation}        
        \begin{aligned}
        { C }_{ L }^{ B }=&\frac { -1 }{ { ( { m }_{ t } }^{ 2 }-{ { m }_{ q } }^{ 2 } ) } \frac { ( a+b ) }{ 16{ \pi  }^{ 2 } } \sum_{ k=0 }^{ 3 }\sum _{ m=1 }^{ 3 }\biggl\{ { { m }_{ t } }^{ 2 } { V }^{ L,k }_{qm}{ { V }^{ L,k }_{tm} }^{ * } \big[B_0({ { m }_{ t } }^{ 2 },{ { m }_{ {\phi}_{kS} } }^{ 2 },{ { m }_{ { q }_{ m }^{ M } } }^{ 2 })\biggr.\\
        &+B_1({ { m }_{ t } }^{ 2 },{ { m }_{ {\phi}_{kS} } }^{ 2 },{ { m }_{ { q }_{ m }^{ M } } }^{ 2 })\big]+{ m }_{ t }{ m }_{ q } { V }^{ R,k }_{qm}{ { V }^{ R,k }_{tm} }^{ * } \big[B_0({ { m }_{ t } }^{ 2 },{ { m }_{ {\phi}_{kS} } }^{ 2 },{ { m }_{ { q }_{ m }^{ M } } }^{ 2 })\\
        &+B_1({ { m }_{ t } }^{ 2 },{ { m }_{ {\phi}_{kS} } }^{ 2 },{ { m }_{ { q }_{ m }^{ M } } }^{ 2 })\big]+{ m }_{ q }{ m }_{ { q }_{ m }^{ M } } { V }^{ R,k }_{qm}{ { V }^{ L,k }_{tm} }^{ * } B_0({ { m }_{ t } }^{ 2 },{ { m }_{ {\phi}_{kS} } }^{ 2 },{ { m }_{ { q }_{ m }^{ M } } }^{ 2 })\\
        &+\biggl.{ m }_{ t }{ m }_{ { q }_{ m }^{ M } } { V }^{ L,k }_{qm}{ { V }^{ R,k }_{tm} }^{ * } B_0({ { m }_{ t } }^{ 2 },{ { m }_{ {\phi}_{kS} } }^{ 2 },{ { m }_{ { q }_{ m }^{ M } } }^{ 2 }) \biggr\} \; ,
        \end{aligned}
\end{equation}
and
\begin{equation}        
        \begin{aligned}
        { C }_{ L }^{ C }=&\frac { -1 }{ 16{ \pi  }^{ 2 } }\sum_{ k=0 }^{ 3 }\sum _{ m=1 }^{ 3 }\biggl\{ (a+b){ V }^{ L,k }_{qm}{ { V }^{ L,k }_{tm} }^{ * } \big[\frac { 1 }{ 2 } -2C_{00}({ { m }_{ t } }^{ 2 },{ { m }_{ Z } }^{ 2 },{ { m }_{ q } }^{ 2 },{ { m }_{ {\phi}_{kS} } }^{ 2 },{ { m }_{ { q }_{ m }^{ M } } }^{ 2 },{ { m }_{ { q }_{ m }^{ M } } }^{ 2 })\big] \biggr.\\
        & +{ { m }_{ { q }_{ m }^{ M } } }^{ 2 } (a-b){ V }^{ L,k }_{qm}{ { V }^{ L,k }_{tm} }^{ * } C_0({ { m }_{ t } }^{ 2 },{ { m }_{ Z } }^{ 2 },{ { m }_{ q } }^{ 2 },{ { m }_{ {\phi}_{kS} } }^{ 2 },{ { m }_{ { q }_{ m }^{ M } } }^{ 2 },{ { m }_{ { q }_{ m }^{ M } } }^{ 2 })\\
        &+{ { m }_{ Z } }^{ 2 } (a+b){ V }^{ L,k }_{qm}{ { V }^{ L,k }_{tm} }^{ * } C_{12}({ { m }_{ t } }^{ 2 },{ { m }_{ Z } }^{ 2 },{ { m }_{ q } }^{ 2 },{ { m }_{ {\phi}_{kS} } }^{ 2 },{ { m }_{ { q }_{ m }^{ M } } }^{ 2 },{ { m }_{ { q }_{ m }^{ M } } }^{ 2 }) \\
        &+{ m }_{ t }{ m }_{ q } (a-b){ V }^{ R,k }_{qm}{ { V }^{ R,k }_{tm} }^{ * } \big[ 2C_1({ { m }_{ t } }^{ 2 },{ { m }_{ Z } }^{ 2 },{ { m }_{ q } }^{ 2 },{ { m }_{ {\phi}_{kS} } }^{ 2 },{ { m }_{ { q }_{ m }^{ M } } }^{ 2 },{ { m }_{ { q }_{ m }^{ M } } }^{ 2 }) \\
        &+2C_2({ { m }_{ t } }^{ 2 },{ { m }_{ Z } }^{ 2 },{ { m }_{ q } }^{ 2 },{ { m }_{ {\phi}_{kS} } }^{ 2 },{ { m }_{ { q }_{ m }^{ M } } }^{ 2 },{ { m }_{ { q }_{ m }^{ M } } }^{ 2 })\\
        &+2C_{12}({ { m }_{ t } }^{ 2 },{ { m }_{ Z } }^{ 2 },{ { m }_{ q } }^{ 2 },{ { m }_{ {\phi}_{kS} } }^{ 2 },{ { m }_{ { q }_{ m }^{ M } } }^{ 2 },{ { m }_{ { q }_{ m }^{ M } } }^{ 2 })\\
        &+C_{11}({ { m }_{ t } }^{ 2 },{ { m }_{ Z } }^{ 2 },{ { m }_{ q } }^{ 2 },{ { m }_{ {\phi}_{kS} } }^{ 2 },{ { m }_{ { q }_{ m }^{ M } } }^{ 2 },{ { m }_{ { q }_{ m }^{ M } } }^{ 2 })\\
        &+C_{22}({ { m }_{ t } }^{ 2 },{ { m }_{ Z } }^{ 2 },{ { m }_{ q } }^{ 2 },{ { m }_{ {\phi}_{kS} } }^{ 2 },{ { m }_{ { q }_{ m }^{ M } } }^{ 2 },{ { m }_{ { q }_{ m }^{ M } } }^{ 2 })\\
        &+ C_0({ { m }_{ t } }^{ 2 },{ { m }_{ Z } }^{ 2 },{ { m }_{ q } }^{ 2 },{ { m }_{ {\phi}_{kS} } }^{ 2 },{ { m }_{ { q }_{ m }^{ M } } }^{ 2 },{ { m }_{ { q }_{ m }^{ M } } }^{ 2 }) \big]\\
        &+{ m }_{ t }{ m }_{ { q }_{ m }^{ M } } (a-b){ V }^{ L,k }_{qm}{ { V }^{ R,k }_{tm} }^{ * } \left[ C_0({ { m }_{ t } }^{ 2 },{ { m }_{ Z } }^{ 2 },{ { m }_{ q } }^{ 2 },{ { m }_{ {\phi}_{kS} } }^{ 2 },{ { m }_{ { q }_{ m }^{ M } } }^{ 2 },{ { m }_{ { q }_{ m }^{ M } } }^{ 2 })\right. \\
        &+C_1({ { m }_{ t } }^{ 2 },{ { m }_{ Z } }^{ 2 },{ { m }_{ q } }^{ 2 },{ { m }_{ {\phi}_{kS} } }^{ 2 },{ { m }_{ { q }_{ m }^{ M } } }^{ 2 },{ { m }_{ { q }_{ m }^{ M } } }^{ 2 })\\
        &+C_2({ { m }_{ t } }^{ 2 },{ { m }_{ Z } }^{ 2 },{ { m }_{ q } }^{ 2 },{ { m }_{ {\phi}_{kS} } }^{ 2 },{ { m }_{ { q }_{ m }^{ M } } }^{ 2 },{ { m }_{ { q }_{ m }^{ M } } }^{ 2 }) \big] \\
        &+{ m }_{ q }{ m }_{ { q }_{ m }^{ M } } (a-b){ V }^{ R,k }_{qm}{ { V }^{ L,k }_{tm} }^{ * } \left[ C_0({ { m }_{ t } }^{ 2 },{ { m }_{ Z } }^{ 2 },{ { m }_{ q } }^{ 2 },{ { m }_{ {\phi}_{kS} } }^{ 2 },{ { m }_{ { q }_{ m }^{ M } } }^{ 2 },{ { m }_{ { q }_{ m }^{ M } } }^{ 2 }) \right. \\
        &+C_1({ { m }_{ t } }^{ 2 },{ { m }_{ Z } }^{ 2 },{ { m }_{ q } }^{ 2 },{ { m }_{ {\phi}_{kS} } }^{ 2 },{ { m }_{ { q }_{ m }^{ M } } }^{ 2 },{ { m }_{ { q }_{ m }^{ M } } }^{ 2 })\\
        &+\biggl. C_2({ { m }_{ t } }^{ 2 },{ { m }_{ Z } }^{ 2 },{ { m }_{ q } }^{ 2 },{ { m }_{ {\phi}_{kS} } }^{ 2 },{ { m }_{ { q }_{ m }^{ M } } }^{ 2 },{ { m }_{ { q }_{ m }^{ M } } }^{ 2 })\big] \biggr\} \; ;
    \end{aligned}
\end{equation}
\begin{equation}
    \begin{aligned}
        { C }_{ R }^{ A }=&\frac { 1 }{ { ( { m }_{ t } }^{ 2 }-{ { m }_{ q } }^{ 2 } ) } \frac { ( a-b ) }{ 16{ \pi  }^{ 2 } } \sum_{ k=0 }^{ 3 }\sum _{ m=1 }^{ 3 }\biggl\{ { { m }_{ q } }^{ 2 } { V }^{ R,k }_{qm}{ { V }^{ R,k }_{tm} }^{ * } \big[ B_0({ { m }_{ q } }^{ 2 },{ { m }_{ {\phi}_{kS} } }^{ 2 },{ { m }_{ { q }_{ m }^{ M } } }^{ 2 }) \biggr.\\
        &+B_1({ { m }_{ q } }^{ 2 },{ { m }_{ {\phi}_{kS} } }^{ 2 },{ { m }_{ { q }_{ m }^{ M } } }^{ 2 }) \big] +{ m }_{ t }{ m }_{ q } { V }^{ L,k }_{qm}{ { V }^{ L,k }_{tm} }^{ * } \big[ B_0({ { m }_{ q } }^{ 2 },{ { m }_{ {\phi}_{kS} } }^{ 2 },{ { m }_{ { q }_{ m }^{ M } } }^{ 2 })\\
        &+B_1({ { m }_{ q } }^{ 2 },{ { m }_{ {\phi}_{kS} } }^{ 2 },{ { m }_{ { q }_{ m }^{ M } } }^{ 2 }) \big] +{ m }_{ q }{ m }_{ { q }_{ m }^{ M } } { V }^{ L,k }_{qm}{ { V }^{ R,k }_{tm} }^{ * } B_0({ { m }_{ q } }^{ 2 },{ { m }_{ {\phi}_{kS} } }^{ 2 },{ { m }_{ { q }_{ m }^{ M } } }^{ 2 })\\
        &+\biggl.{ m }_{ t }{ m }_{ { q }_{ m }^{ M } } { V }^{ R,k }_{qm}{ { V }^{ L,k }_{tm} }^{ * }  B_0({ { m }_{ q } }^{ 2 },{ { m }_{ {\phi}_{kS} } }^{ 2 },{ { m }_{ { q }_{ m }^{ M } } }^{ 2 })\biggr\} \; ,
        \end{aligned}
\end{equation}
\begin{equation}        
        \begin{aligned}        
        { C }_{ R }^{ B}=&\frac { -1 }{ { ( { m }_{ t } }^{ 2 }-{ { m }_{ q } }^{ 2 }) } \frac { ( a-b ) }{ 16{ \pi  }^{ 2 } }\sum_{ k=0 }^{ 3 }\sum _{ m=1 }^{ 3 }\biggl\{ { { m }_{ t } }^{ 2 } { V }^{ R,k }_{qm}{ { V }^{ R,k }_{tm} }^{ * } \big[B_0({ { m }_{ t } }^{ 2 },{ { m }_{ {\phi}_{kS} } }^{ 2 },{ { m }_{ { q }_{ m }^{ M } } }^{ 2 })\biggr.\\
        &+B_1({ { m }_{ t } }^{ 2 },{ { m }_{ {\phi}_{kS} } }^{ 2 },{ { m }_{ { q }_{ m }^{ M } } }^{ 2 }) \big]+{ m }_{ t }{ m }_{ q } { V }^{ L,k }_{qm}{ { V }^{ L,k }_{tm} }^{ * } \big[B_0({ { m }_{ t } }^{ 2 },{ { m }_{ {\phi}_{kS} } }^{ 2 },{ { m }_{ { q }_{ m }^{ M } } }^{ 2 })\\
        &+B_1({ { m }_{ t } }^{ 2 },{ { m }_{ {\phi}_{kS} } }^{ 2 },{ { m }_{ { q }_{ m }^{ M } } }^{ 2 }) \big]+{ m }_{ q }{ m }_{ { q }_{ m }^{ M } }{ V }^{ L,k }_{qm}{ { V }^{ R,k }_{tm} }^{ * } B_0({ { m }_{ t } }^{ 2 },{ { m }_{ {\phi}_{kS} } }^{ 2 },{ { m }_{ { q }_{ m }^{ M } } }^{ 2 })\\
        &+\biggl.{ m }_{ t }{ m }_{ { q }_{ m }^{ M } } { V }^{ R,k }_{qm}{ { V }^{ L,k }_{tm} }^{ * } B_0({ { m }_{ t } }^{ 2 },{ { m }_{ {\phi}_{kS} } }^{ 2 },{ { m }_{ { q }_{ m }^{ M } } }^{ 2 })\biggr\} \; ,
        \end{aligned}
\end{equation}
and
\begin{equation}        
        \begin{aligned}        
        { C }_{ R }^{ C }=&\frac { -1 }{ 16{ \pi  }^{ 2 } }\sum_{ k=0 }^{ 3 }\sum _{ m=1 }^{ 3 }\biggl\{ (a-b){ V }^{ R,k }_{qm}{ { V }^{ R,k }_{tm} }^{ * } \big[\frac { 1 }{ 2 } -2C_{00}({ { m }_{ t } }^{ 2 },{ { m }_{ Z } }^{ 2 },{ { m }_{ q } }^{ 2 },{ { m }_{ {\phi}_{kS} } }^{ 2 },{ { m }_{ { q }_{ m }^{ M } } }^{ 2 },{ { m }_{ { q }_{ m }^{ M } } }^{ 2 })\big]\biggr.\\
        &+{ { m }_{ { q }_{ m }^{ M } } }^{ 2 } (a+b){ V }^{ R,k }_{qm}{ { V }^{ R,k }_{tm} }^{ * } C_0({ { m }_{ t } }^{ 2 },{ { m }_{ Z } }^{ 2 },{ { m }_{ q } }^{ 2 },{ { m }_{ {\phi}_{kS} } }^{ 2 },{ { m }_{ { q }_{ m }^{ M } } }^{ 2 },{ { m }_{ { q }_{ m }^{ M } } }^{ 2 })\\
        &+{ { m }_{ Z } }^{ 2 } (a-b){ V }^{ R,k }_{qm}{ { V }^{ R,k }_{tm} }^{ * }  C_{12}({ { m }_{ t } }^{ 2 },{ { m }_{ Z } }^{ 2 },{ { m }_{ q } }^{ 2 },{ { m }_{ {\phi}_{kS} } }^{ 2 },{ { m }_{ { q }_{ m }^{ M } } }^{ 2 },{ { m }_{ { q }_{ m }^{ M } } }^{ 2 })\\
        &+{ m }_{ t }{ m }_{ q }(a+b){ V }^{ L,k }_{qm}{ { V }^{ L,k }_{tm} }^{ * } \big[ 2C_1({ { m }_{ t } }^{ 2 },{ { m }_{ Z } }^{ 2 },{ { m }_{ q } }^{ 2 },{ { m }_{ {\phi}_{kS} } }^{ 2 },{ { m }_{ { q }_{ m }^{ M } } }^{ 2 },{ { m }_{ { q }_{ m }^{ M } } }^{ 2 })\\
        &+2C_2({ { m }_{ t } }^{ 2 },{ { m }_{ Z } }^{ 2 },{ { m }_{ q } }^{ 2 },{ { m }_{ {\phi}_{kS} } }^{ 2 },{ { m }_{ { q }_{ m }^{ M } } }^{ 2 },{ { m }_{ { q }_{ m }^{ M } } }^{ 2 })\\
        &+2C_{12}({ { m }_{ t } }^{ 2 },{ { m }_{ Z } }^{ 2 },{ { m }_{ q } }^{ 2 },{ { m }_{ {\phi}_{kS} } }^{ 2 },{ { m }_{ { q }_{ m }^{ M } } }^{ 2 },{ { m }_{ { q }_{ m }^{ M } } }^{ 2 })\\
        &+C_{11}({ { m }_{ t } }^{ 2 },{ { m }_{ Z } }^{ 2 },{ { m }_{ q } }^{ 2 },{ { m }_{ {\phi}_{kS} } }^{ 2 },{ { m }_{ { q }_{ m }^{ M } } }^{ 2 },{ { m }_{ { q }_{ m }^{ M } } }^{ 2 })\\
        &+C_{22}({ { m }_{ t } }^{ 2 },{ { m }_{ Z } }^{ 2 },{ { m }_{ q } }^{ 2 },{ { m }_{ {\phi}_{kS} } }^{ 2 },{ { m }_{ { q }_{ m }^{ M } } }^{ 2 },{ { m }_{ { q }_{ m }^{ M } } }^{ 2 })\\
        &+C_0({ { m }_{ t } }^{ 2 },{ { m }_{ Z } }^{ 2 },{ { m }_{ q } }^{ 2 },{ { m }_{ {\phi}_{kS} } }^{ 2 },{ { m }_{ { q }_{ m }^{ M } } }^{ 2 },{ { m }_{ { q }_{ m }^{ M } } }^{ 2 }) \big] \\
        &+{ m }_{ t }{ m }_{ { q }_{ m }^{ M } }(a+b){ V }^{ R,k }_{qm}{ { V }^{ L,k }_{tm} }^{ * } \big[ C_0({ { m }_{ t } }^{ 2 },{ { m }_{ Z } }^{ 2 },{ { m }_{ q } }^{ 2 },{ { m }_{ {\phi}_{kS} } }^{ 2 },{ { m }_{ { q }_{ m }^{ M } } }^{ 2 },{ { m }_{ { q }_{ m }^{ M } } }^{ 2 })\\
        &+C_1({ { m }_{ t } }^{ 2 },{ { m }_{ Z } }^{ 2 },{ { m }_{ q } }^{ 2 },{ { m }_{ {\phi}_{kS} } }^{ 2 },{ { m }_{ { q }_{ m }^{ M } } }^{ 2 },{ { m }_{ { q }_{ m }^{ M } } }^{ 2 })\\
        &+C_2({ { m }_{ t } }^{ 2 },{ { m }_{ Z } }^{ 2 },{ { m }_{ q } }^{ 2 },{ { m }_{ {\phi}_{kS} } }^{ 2 },{ { m }_{ { q }_{ m }^{ M } } }^{ 2 },{ { m }_{ { q }_{ m }^{ M } } }^{ 2 }) \big] \\
        &+{ m }_{ q }{ m }_{ { q }_{ m }^{ M } } (a+b){ V }^{ L,k }_{qm}{ { V }^{ R,k }_{tm} }^{ * } \big[ C_0({ { m }_{ t } }^{ 2 },{ { m }_{ Z } }^{ 2 },{ { m }_{ q } }^{ 2 },{ { m }_{ {\phi}_{kS} } }^{ 2 },{ { m }_{ { q }_{ m }^{ M } } }^{ 2 },{ { m }_{ { q }_{ m }^{ M } } }^{ 2 })\\
        &+C_1({ { m }_{ t } }^{ 2 },{ { m }_{ Z } }^{ 2 },{ { m }_{ q } }^{ 2 },{ { m }_{ {\phi}_{kS} } }^{ 2 },{ { m }_{ { q }_{ m }^{ M } } }^{ 2 },{ { m }_{ { q }_{ m }^{ M } } }^{ 2 })\\
        &+\biggl.C_2({ { m }_{ t } }^{ 2 },{ { m }_{ Z } }^{ 2 },{ { m }_{ q } }^{ 2 },{ { m }_{ {\phi}_{kS} } }^{ 2 },{ { m }_{ { q }_{ m }^{ M } } }^{ 2 },{ { m }_{ { q }_{ m }^{ M } } }^{ 2 })
\big] \biggr\} \; .
    \end{aligned}
\end{equation}
Each of the above contributions $C^A_{L,R}$, $C^B_{L,R}$ and $C^C_{L,R}$ 
are ultraviolet divergent. However by using the divergent parts of the Passarino-Veltman (PV) functions
\begin{equation}
\begin{aligned}
{\rm Div}[B_0] & = +\Delta_\epsilon  \; ,\\
{\rm Div}[B_1] & = -\frac{1}{2}\Delta_\epsilon \; ,\\
{\rm Div}[C_{00}] & = +\frac{1}{4}\Delta_\epsilon \; ,
\end{aligned}
\end{equation}
where $\Delta_\epsilon = 2/\epsilon - \gamma_E  + \ln 4 \pi$ 
with $\epsilon = 4-d$ is the regulator in dimensional regularization 
and $\gamma_E$ being the Euler's constant, one can easily verify that the divergences in the three diagrams
summed up to nil leading to finite results for $C_L$ and $C_R$.

Only Diagram (C) contributes to the dipole form factors $A_L$  and  $A_R$. 
They are given by
\begin{equation}
    \begin{aligned}
        { A }_{ L }=&\frac { 1 }{ 16{ \pi  }^{ 2 } }\sum_{ k=0 }^{ 3 }\sum _{ m=1 }^{ 3 }\biggl\{ \frac { { { m }_{ t } }^{ 2 } }{ 2 } (a-b){ V }^{ R,k }_{qm}{ { V }^{ R,k }_{tm} }^{ * }\biggl[ C_1({ { m }_{ t } }^{ 2 },{ { m }_{ Z } }^{ 2 },{ { m }_{ q } }^{ 2 },{ { m }_{ {\phi}_{kS} } }^{ 2 },{ { m }_{ { q }_{ m }^{ M } } }^{ 2 },{ { m }_{ { q }_{ m }^{ M } } }^{ 2 })\biggr. \biggr.\\
        &+C_{11}({ { m }_{ t } }^{ 2 },{ { m }_{ Z } }^{ 2 },{ { m }_{ q } }^{ 2 },{ { m }_{ {\phi}_{kS} } }^{ 2 },{ { m }_{ { q }_{ m }^{ M } } }^{ 2 },{ { m }_{ { q }_{ m }^{ M } } }^{ 2 })+\biggl. C_{12}({ { m }_{ t } }^{ 2 },{ { m }_{ Z } }^{ 2 },{ { m }_{ q } }^{ 2 },{ { m }_{ {\phi}_{kS} } }^{ 2 },{ { m }_{ { q }_{ m }^{ M } } }^{ 2 },{ { m }_{ { q }_{ m }^{ M } } }^{ 2 })\biggr]\\
        &+\frac { { m }_{ t }{ m }_{ q } }{ 2 } (a+b){ V }^{ L,k }_{qm}{ { V }^{ L,k }_{tm} }^{ * } \biggl[ C_2({ { m }_{ t } }^{ 2 },{ { m }_{ Z } }^{ 2 },{ { m }_{ q } }^{ 2 },{ { m }_{ {\phi}_{kS} } }^{ 2 },{ { m }_{ { q }_{ m }^{ M } } }^{ 2 },{ { m }_{ { q }_{ m }^{ M } } }^{ 2 }) \biggr. \\
        &+C_{12}({ { m }_{ t } }^{ 2 },{ { m }_{ Z } }^{ 2 },{ { m }_{ q } }^{ 2 },{ { m }_{ {\phi}_{kS} } }^{ 2 },{ { m }_{ { q }_{ m }^{ M } } }^{ 2 },{ { m }_{ { q }_{ m }^{ M } } }^{ 2 })+\biggl. C_{22}({ { m }_{ t } }^{ 2 },{ { m }_{ Z } }^{ 2 },{ { m }_{ q } }^{ 2 },{ { m }_{ {\phi}_{kS} } }^{ 2 },{ { m }_{ { q }_{ m }^{ M } } }^{ 2 },{ { m }_{ { q }_{ m }^{ M } } }^{ 2 })\biggr]\\
        &+\frac { { m }_{ t }{ m }_{ { q }_{ m }^{ M } } }{ 2 } { V }^{ R,k }_{qm}{ { V }^{ L,k }_{tm} }^{ * } 
        \biggl[ (a-b)C_1({ { m }_{ t } }^{ 2 },{ { m }_{ Z } }^{ 2 },{ { m }_{ q } }^{ 2 },{ { m }_{ {\phi}_{kS} } }^{ 2 },{ { m }_{ { q }_{ m }^{ M } } }^{ 2 },{ { m }_{ { q }_{ m }^{ M } } }^{ 2 }) \biggr. \\
        &\biggl. \biggl. +(a+b)C_2({ { m }_{ t } }^{ 2 },{ { m }_{ Z } }^{ 2 },{ { m }_{ q } }^{ 2 },{ { m }_{ {\phi}_{kS} } }^{ 2 },{ { m }_{ { q }_{ m }^{ M } } }^{ 2 },{ { m }_{ { q }_{ m }^{ M } } }^{ 2 })\biggr] \biggr\} \; ,
        \end{aligned}
\end{equation}
and
\begin{equation}        
        \begin{aligned}        
        { A }_{ R }=&\frac { 1 }{ 16{ \pi  }^{ 2 } }\sum_{ k=0 }^{ 3 }\sum _{ m=1 }^{ 3 }\biggl\{ \frac { { { m }_{ t } }^{ 2 } }{ 2 } (a+b){ V }^{ L,k }_{qm}{ { V }^{ L,k }_{tm} }^{ * }\biggl[ C_1({ { m }_{ t } }^{ 2 },{ { m }_{ Z } }^{ 2 },{ { m }_{ q } }^{ 2 },{ { m }_{ {\phi}_{kS} } }^{ 2 },{ { m }_{ { q }_{ m }^{ M } } }^{ 2 },{ { m }_{ { q }_{ m }^{ M } } }^{ 2 }) \biggr. \biggr. \\
        &+C_{11}({ { m }_{ t } }^{ 2 },{ { m }_{ Z } }^{ 2 },{ { m }_{ q } }^{ 2 },{ { m }_{ {\phi}_{kS} } }^{ 2 },{ { m }_{ { q }_{ m }^{ M } } }^{ 2 },{ { m }_{ { q }_{ m }^{ M } } }^{ 2 })+\biggl. C_{12}({ { m }_{ t } }^{ 2 },{ { m }_{ Z } }^{ 2 },{ { m }_{ q } }^{ 2 },{ { m }_{ {\phi}_{kS} } }^{ 2 },{ { m }_{ { q }_{ m }^{ M } } }^{ 2 },{ { m }_{ { q }_{ m }^{ M } } }^{ 2 })\biggr]\\
        &+\frac { { m }_{ t }{ m }_{ q } }{ 2 } (a-b){ V }^{ R,k }_{qm}{ { V }^{ R,k }_{tm} }^{ * } \biggl[ C_2({ { m }_{ t } }^{ 2 },{ { m }_{ Z } }^{ 2 },{ { m }_{ q } }^{ 2 },{ { m }_{ {\phi}_{kS} } }^{ 2 },{ { m }_{ { q }_{ m }^{ M } } }^{ 2 },{ { m }_{ { q }_{ m }^{ M } } }^{ 2 }) \biggr. \\
        &+C_{12}({ { m }_{ t } }^{ 2 },{ { m }_{ Z } }^{ 2 },{ { m }_{ q } }^{ 2 },{ { m }_{ {\phi}_{kS} } }^{ 2 },{ { m }_{ { q }_{ m }^{ M } } }^{ 2 },{ { m }_{ { q }_{ m }^{ M } } }^{ 2 })+\biggl. C_{22}({ { m }_{ t } }^{ 2 },{ { m }_{ Z } }^{ 2 },{ { m }_{ q } }^{ 2 },{ { m }_{ {\phi}_{kS} } }^{ 2 },{ { m }_{ { q }_{ m }^{ M } } }^{ 2 },{ { m }_{ { q }_{ m }^{ M } } }^{ 2 })\biggr]\\
        &+\frac { { m }_{ t }{ m }_{ { q }_{ m }^{ M } } }{ 2 } { V }^{ L,k }_{qm}{ { V }^{ R,k }_{tm} }^{ * }
        \biggl[ (a+b)C_1({ { m }_{ t } }^{ 2 },{ { m }_{ Z } }^{ 2 },{ { m }_{ q } }^{ 2 },{ { m }_{ {\phi}_{kS} } }^{ 2 },{ { m }_{ { q }_{ m }^{ M } } }^{ 2 },{ { m }_{ { q }_{ m }^{ M } } }^{ 2 }) \biggr. \\
        &+ \biggl. \biggl. (a-b)C_2({ { m }_{ t } }^{ 2 },{ { m }_{ Z } }^{ 2 },{ { m }_{ q } }^{ 2 },{ { m }_{ {\phi}_{kS} } }^{ 2 },{ { m }_{ { q }_{ m }^{ M } } }^{ 2 },{ { m }_{ { q }_{ m }^{ M } } }^{ 2 })\biggr] \biggr\} \; .
    \end{aligned}
\end{equation}
Since the PV functions $C_1,C_2,C_{11},C_{12}$ and $C_{22}$ do not have 
ultraviolet divergences, $A_L$ and $A_R$ are finite, as one should expect for they 
are the coefficients of the non-renormalizable magnetic and electric dipole operators.

\subsection*{Form Factors for $t \to \gamma q$ and $t \to g q$}

$A'_L$ and $A'_R$ can be obtained from the above $A_L$ and $A_R$ respectively 
by replacing
\begin{equation}
m_Z^2  \rightarrow  0 \;  , \;\;\;\;
a  \rightarrow  \frac{2}{3}e  \; , \;\;\;\;
b  \rightarrow 0 \; .
\end{equation}

The decay rate for $t \to g q$ can be obtained from that of
$t \to \gamma q$ simply by replacing 
the top quark electric charge $\frac{2}{3}e$ 
by the strong coupling $g_s$
and multiply the final result by an overall color factor 
$(N_C^2-1)/2N_C$ where $N_C$ is the number of color.
Thus
\begin{equation}
\frac{\Gamma (t \to g q)}{\Gamma (t \to \gamma q)} = \frac{9}{4} \cdot \frac{N_C^2 - 1}{2 N_C} \cdot \frac{\alpha_s}{\alpha_{\rm em}}.
\end{equation}
Taking $N_C=3$, $\alpha_s = 0.11$ and $\alpha^{-1}_{\rm em} = 128$, this ratio is about 42.
Next-to-leading order QCD corrections to the processes $t\to\gamma q$, $t \to Z q$ and $t \to g q$ can be found in 
\cite{Zhang:2008yn,Zhang:2010bm}.
Moreover, the next-to-leading order and next-to-next-to-leading order QCD corrections for the dominant SM top quark decay 
mode $t \to W^+b$ had  been computed in~\cite{Jezabek:1988iv,Li:1990qf} and
\cite{Gao:2012ja} respectively.



\end{document}